\documentclass[fleqn,usenatbib]{mnras}
 \usepackage{xr-hyper}

\usepackage{hyperref}	
\hypersetup{colorlinks=true,linkcolor=blue,citecolor=blue,filecolor=blue,urlcolor=blue}

\usepackage{newtxtext,newtxmath}

\usepackage[utf8]{inputenc}
\usepackage[T1]{fontenc}
\usepackage{ae,aecompl}
\usepackage{graphicx}	
\usepackage{multicol}        
\usepackage{bm}         
\usepackage{pdflscape}  
\usepackage{enumitem}
\usepackage{xspace}
\usepackage{hhline}

\graphicspath{{./figures/}}

\usepackage{color}

\newcommand{\vdva}{} 
\newcommand{\vtri}{} 
\newcommand{\vfour}{} 
 
\newcommand{\vsix}{} 
\newcommand{\vseven}{}

\title[Disc evolution around magnetized star]{Physical modeling of viscous disc evolution around magnetized neutron star. Aql X-1 2013 outburst decay}
\author[G. Lipunova et al.]
{Galina Lipunova,$^{1}$\thanks{E-mail: gvlipunova@sai.msu.ru}    
Konstantin Malanchev,$^{1,2}$ 
Sergey Tsygankov,$^{3,4}$ 
Nikolai Shakura,$^{1,5}$ 
\newauthor
Andrei Tavleev, $^{6,1}$
Dmitry Kolesnikov $^{1}$
\\
$^{1}$Moscow Lomonosov State University Sternberg Astronomical Institute, Moscow 119992, Universitetskiy pr., 13, Russia\\
$^{2}$Department of Astronomy, University of Illinois at Urbana-Champaign, 1002 West Green Street, Urbana, IL 61801, USA\\
$^{3}$Department of Physics and Astronomy,  FI-20014 University of Turku, Finland \\ 
$^{4}$Space Research Institute of the Russian Academy of Sciences, Profsoyuznaya Str. 84/32, Moscow 117997, Russia \\
$^{5}$Kazan Federal University, 420008 Kazan, Russia\\
$^{6}$Physics Faculty of Moscow Lomonosov State University, Moscow 119992, Russia
}

\date{Accepted 2021 November 10. Received 2021 November 10; in original form 2021 March 5}

\pubyear{2022}
 \makeatletter
 \newcommand*{\addFileDependency}[1]{
   \typeout{(#1)}
   \@addtofilelist{#1}
   \IfFileExists{#1}{}{\typeout{No file #1.}}
 }
 \makeatother
 
 \newcommand*{\myexternaldocument}[1]{%
     \externaldocument{#1}%
     \addFileDependency{#1.tex}%
     \addFileDependency{#1.aux}%
 }
\myexternaldocument{aux/pdf_figures}

\begin{document}
\label{firstpage}
\pagerange{\pageref{firstpage}--\pageref{lastpage}}
\maketitle

\def\dMed{\dot M_\mathrm{Edd}}
\def\dMin{\dot M_\mathrm{in}}
\def\dMout{\dot M_\mathrm{out}}
\def\Fin{F_\mathrm{in}}
\def\Fd{F_\mathrm{dead}}
\def\Fm{F_\mathrm{m}}
\def\hout{h_\mathrm{hot}}
\def\Ledd{L_\mathrm{Edd}}
\def\ledd{L_\mathrm{Edd}}
\def\mag{\mathrm{mag}}
\def\Rc{R_\mathrm{cor}}
\def\Rg{R_\mathrm{s}}  
\def\rg{r_\mathrm{g}}  
\def\Rd{R_\mathrm{dead}}
\def\Rdead{R_\mathrm{dead}}
\def\Rmso{R_\mathrm{mso}}
\def\Rcor{R_\mathrm{cor}}
\def\Rin{R_\mathrm{in}}
\def\Rhot{R_\mathrm{hot}}
\def\Rm{R_\mathrm{mag}}
\def\Rms{R_\mathrm{ms}}
\def\sigmat{\sigma_{\scriptsize\mbox{\textsc{t}}}}
\def\xinner{\xi_\mathrm{mag}}
\def\v{\textit{v}}
\newcommand{\Chot}{\widetilde{\mathcal{C}}_{\rm irr}\xspace}
\newcommand{\Cirr}{{\mathcal{C}}_{\rm irr}\xspace}
\newcommand{\Ccold}{\widetilde{\mathcal{C}}^{\rm cold}_{\rm irr}\xspace}
\newcommand{\cod}{\mathcal{R}^2\xspace}
\newcommand{\eff}{\mathrm{eff}}
\newcommand{\hin}{h_\mathrm{in}}
\newcommand{\freddi}{{\sc freddi}\xspace}
\newcommand{\Fvis}{F_\mathrm{vis}}
\newcommand{\Fmag}{F_\mathrm{mag}}
\newcommand{\irr}{\mathrm{irr}}
\newcommand{\mns}{M_{\star}}
\newcommand{\Msun}{M_\odot}
\newcommand{\ksu}{K_\mathrm{su}}
\newcommand{\ksd}{K_\mathrm{sd}}
\newcommand{\kappat}{\kappa_\mathrm{t}}
\newcommand{\kappatd}{\kappa_\mathrm{td}}
\newcommand{\NS}{SA\xspace}
\newcommand{\PONS}{POSA\xspace}
\newcommand{\PO}{PO\xspace}
\newcommand{\rc}{R_{\mathrm{cor}}}
\newcommand{\rin}{R_{\mathrm{in}}}
\newcommand{\rscr}{R_\mathrm{scr}}
\newcommand{\scr}{\mathrm{scr}}
\newcommand{\vis}{\mathrm{vis}}
\newcommand{\Wrf}{W_{r\varphi}}
\maketitle
\def\SAmain{SA1\xspace}
\def\SAmainSmallAlpha{SA2\xspace}
\def\SAmaink0{SA3\xspace}
\def\SAmainNoCold{SA4\xspace}
\def\SAmainNoColdStar{SA5\xspace}
\def\SAmainNoOpt{SA6\xspace}
\def\smallB{SA7\xspace}
\def\smallBsmalla{SA8\xspace}
\def\POSAmain{POSA9\xspace}
\def\POmain{PO10\xspace}
\def\SARcor{SA11\xspace}
\def\SARdead{SA12\xspace}
\def\irrI{PO13\xspace}
\def\irrII{PO14\xspace}
\def\irrIII{PO15\xspace}
\def\irrIV{PO16\xspace}
\def\irrX{PO19\xspace}

\begin{abstract}
We present a model of a viscously evolving accretion disc around a magnetized neutron star.
The model features the varying outer radius of the hot ionized part of the disc due to cooling and the varying inner radius of the disc due to interaction with the magnetosphere. It also includes hindering of accretion on the neutron star because of the centrifugal barrier and  irradiation of the outer disc and  companion star by  X-rays from the neutron star and disc. 
When setting inner boundary conditions, we take into account that processes at the inner disc occur on a time scale much less than the viscous time scale of the whole disc.
{\vsix We consider three types of outflow from the disc inner edge: zero outflow, one based on MHD calculations,  and a very efficient propeller mechanism.}
The light curves of an X-ray transient after the outburst peak can be calculated by a corresponding, publicly available code. 
We compare observed light curves of the 2013 burst of  Aql\,X-1  in  X-ray and optical bands with  modeled ones.
We find that the fast drop of the $0.3-10$~keV flux can be solely explained by a radial shrinking of the hot disc. At the same time, models with the neutron star magnetic field $>10^8$~G have better fits because the accretion efficiency behaviour emphasizes the 'knee' on the light curve.
We also find that a plato emission can be produced by a disc-reservoir with stalled accretion.
\end{abstract}

\section{Introduction}

Brightest transient phenomena in  X-ray binary systems are closely connected to long-term dynamical evolution of viscous accretion discs around
compact objects. {\vsix A change of the mass accretion rate on a compact object, which is observed as an outburst lasting many days, can be induced by an instability in a disc or a rise of mass income from a companion star. 
In low-mass X-ray binaries (LMXBs), for instance, the neighbour star fills its Roche lobe and leaks  slowly into the disc until an outburst occurs.
     The thermal–viscous disc
instability~\citep[see, e.g.,][and references therein]{Hameury2020} is now generally thought to be the basic cause of outbursts of
dwarf novae and low-mass X-ray binaries (LMXBs). According to this model, ionization wave goes through a substantial part of a disc before a burst reaches its peak. The temperature and viscosity coefficient in the ionized zone rise. Near the peak the size of the ionized hot zone is the largest and the disc radial structure can be described by  a quasi-stationary, standard $\alpha$-disc.}

Viscous evolution of an accretion disc can be described by a single equation, which follows from equations of the mass and angular momentum conservation. 
This is true when assumptions underlying the standard disc model~\citep{shakura-sunyaev1973} hold: a disc is geometrically thin and optically thick and the local energy balance works.
The standard disc model is widely and successfully used to interpret observations and to describe disc physics. 
Plugged into the viscous evolution equation, the standard disc with constant outer radius produces a fast-rise  quasi-exponential-decay  (FRED) light curve~\citep{lip-sha2000,lipunova2015}.
However, FRED light curves are rather exceptional cases. Observed irregularities, platoes, breaks, {\vsix and reflares} of the light curves  demonstrate physical complexity calling for further analytic and numerical considerations~\citep[see, e.g.,][]{Baginska+2021}. 
  
Comparing discs around neutrons stars and black holes, we bound to acknowledge that disc evolution around neutron stars is even more complicated and fascinating. 
The neutron stars are extreme objects and emitters of the  electromagnetic radiation of various kinds~\citep[e.g.,][]{Ozel-Freire2016}. 
Different species of sources with neutron stars are observed thanks to the various combination of the key parameters: the spin and magnetic field of the star, the density and angular momentum of surrounding matter~\citep{lipunov1992}. 

In X-ray transients, both with black holes (BHXT) and neutron stars (NSXT), neat FRED light curves are exceptional.  
A common feature is a change of the light curve slope during a decay {\vsix to a faster one.}  Models, explaining such non-trivial light curves of BHXTs subsist exclusively on the developments in the discs themselves.
In the NSXTs,  the physics of the central star may have impact on the light curves. Sudden drops of X-ray flux are sometimes regarded as  manifestations of the accretion rate being blocked on its way to the rotating neutron star by a sufficiently large magnetosphere~\citep{Stella+1986,Cui1997,Campana+1998, Gilfanov+1998, Raguzova-Lipunov1998, zhang+1998b,Alpar2001,Hartman+2009,Hartman+2011, asai+2013,matsuoka-asai2013,Campana+2014,Tsygankov+2016, Furst_etal2017_propeller, Lutovinov+2017}.
The hindering of accretion onto a neutron star can occur  if the disc is disrupted by the magnetic field beyond the corotatation radius~\citep{Shvartsman1970,pringle-rees1972,Lamb+1973,davidson-ostriker1973}, defined as the radius where the Keplerian frequency equals the star's spin frequency.
In such situation, as suggested by theoretical and numerical modelings, the matter can be expelled from the system, due to a `propeller effect'~\citep{Illarionov-Sunyaev1975}, and/or remain in the disc~\citep[][hereafter R18]{Sunyaev-Shakura1977,lipunov1980,Spruit-Taam1993,dangelo-spruit2010,Hartman+2011, dangelo-spruit2012,zanni-ferreira2013,Parfrey-Tchekh2017, parfrey+2017,Romanova+2018}.
 
\defcitealias{Romanova+2018}{R18}

On the other hand, a specific change in the disc evolution, manifested as a steepening of a light curve slope, can happen due to a critical cooling at the outer disc, which leads to a  decrease of the hot disc size\footnote{Usually this change is called a `transition from an exponential decay to a linear one' since such laws of $\dot M(t)$ are realized in the models with constant viscosity.}~\citep{kin-rit1998, Gilfanov+1998,Shahbaz+1998,Powell+2007,Campana+2013}. 
This is an essential component of the `DIM' --- a Disc Instability Model. 
Then, theoretically, the propeller effect is not required to explain a fast drop of the flux.  
A picture unifying the DIM scenario and the propeller effect was considered by \citet{Hartman+2011,Gungor+2014}.

In the present work we take a step towards a detailed modeling of a burst in accretion discs of NSXTs. 
To address a complex evolution of an accretion disc around a magnetized neutron star, we have built a multi-component physical 1D radial model to calculate the disc evolution after the peak of an outburst.
Up to now, numerical 3D simulations can hardly address the whole disc.
This proceeds from the vast difference of scales, temporal and spatial, which need to be involved in a resource-consuming 3D model of a whole disc.  
The 1D studies {\vsix of accretion on neutron stars} also tend to focus on the innermost disc radii~\citep{Spruit-Taam1993,dangelo-spruit2010} with an exception of the work by  \citet{armitage-clarke96} who studied non-steady accretion discs around single and binary magnetized T Tauri stars.

{\vfour  Our model includes  principal ingredients of the evolution of an accretion disc around a neutron  star.}
We do not attempt a deep sophistication of each ingredient of our model at this stage. 
Instead, proper parametrizations are proposed that allow us to construct a reasonably fast numerical scheme which calculates light curves of an accreting neutron star surrounded by a viscously-evolving accretion disc. 
A corresponding  numerical computer code\footnote{available at GitHub through \url{http://xray.sai.msu.ru/sciwork}}  is based on a previously published \freddi-code, which calculates the burst evolution of a BHXT~\citep{Malanchev-Lipunova2016,lipunova-malanchev2017}.

We demonstrate the capacity of our code by successfully modeling an outburst of a well-studied X-ray transient -- \hbox{Aql\,X-1}. 
In 2013, its outburst was observed by {\vsix X-ray instruments, including the Neil Gehrels Swift observatory,} and also by many optical ground observatories. 
The characteristic knee on the light curve was observed around $\sim 45$~day from the peak. 
Using our numerical code, we successfully explain such evolution of flux in X-ray and optical.

We find that to explain the X-ray light curve of Aql\,X-1 of the 2013 burst,  the size of the zone with ionized material   should be decreasing. 
This is in agreement with the DIM model. 
We study if the observed light curves demonstrate  any indications of the magnetosphere-disc interaction {\vsix within our scheme.}
We also explore the possibility  
that the plato X-ray emission is generated by the heat released in a remnant disc with the inner edge stopped beyond the corotation radius.

In \S\ref{s.model}, we present our model. 
In \S\ref{s.data}, the observational data and its reduction are described. Results of modeling can be found in \S\ref{s.results}. 
We compare our model and its results with others, discuss low plato emission and the effect of the neutron star magnetic field and  irradiation parameter's value on the light curves in~\S\ref{s.disc}.
We summarize in \S\ref{s.summmary}.

\section{Model of magnetically-truncated  self-irradiated disc during outburst}\label{s.model}

Viscous evolution of an accretion disc around a central object of mass $M_\star$ is described by the equation of diffusion type~\citep[see, e.g.,][]{lyub-shak1987}:
\begin{equation}
    \frac{ \partial \Sigma }{\partial t} = \frac1{4\pi} \frac{(G \mns)^2}{h^3} 
\frac{\partial^2 F}{\partial h^2},
\label{eq.diffusion}
\end{equation}
where $h \equiv \sqrt{G \mns r}$ is the specific angular momentum, $\Sigma$ is the surface density, $F=2\,\pi\,W_{r\varphi}\,r^2$ is the viscous torque related to the height-integrated viscous stress tensor. 
For the $\alpha$-disc~\citep{shakura1972}, the latter is expressed as
\begin{equation}
   W_{r\varphi} =  \int\limits_{-z_0}^{+z_0}\alpha  P \, {\rm d}z\, , 
\label{eq.Wrfi}
\end{equation}
where $P$ is the pressure in the disc, $z_0$ is the semithickness.

The accretion rate in an evolving disc can be expressed as follows: \begin{equation}
\dot M = \frac{\partial F}{\partial h}\, .
\label{eq:Mdot}
\end{equation}

Equation \eqref{eq.diffusion} can be solved analytically in a number of cases~\citep[][]{lyn-pri1974, lyub-shak1987,pringle1991,tanaka2011,Eksi2012,Illenseer-Duschl2015,lipunova2015,rafikov2016,Balbus-Mummery2018,Mushtukov+2019,Nixon-Pringle2020}. 
Such solutions imply uniform properties of viscosity and uniform analytic relation $F(\Sigma)$ over a disc and its fixed or freely expanding outer radius.
In physically motivated models of discs, realistic constrains complicate the disc evolution calling for numerical approaches. 
First, the boundaries of the fast-evolving hot zone of the disc move.
Second, a numerical (non-analytic) solution of the vertical structure can provide a  relation between the surface density $\Sigma$ and the viscous torque $F$.

Let us emphasize some key features demonstrated by analytic solutions for the discs with limited outer radius.
It is established that the rate of the disc mass variation is determined by the {\em values} at the disc outer radius and the {\em type} of condition at the inner radius.  
The inner rings of the disc adjust themselves according to an inner boundary condition on the local viscous time scale, that is, significantly faster than the global viscous time scale $t_{\rm visc}$.
This means that in a non-stationary disc with  time-dependent $\dot{M}(t)$ but a fixed inner-boundary condition, a quasi-stationary distribution of parameters, determined by the inner condition, is applicable in its inner part.  
If the matter freely leaves the disc at its inner edge then {\vsix the accretion rate is approximately constant over radius} at each moment, $\dot M(r) \approx const$,  and the disc mass varies as  $\dot M_{\rm disc} \sim  M_{\rm disc} /t_{\rm visc} \propto F_{\rm out}/h_{\rm out}$; if, on the contrary, the mass flow through the inner radius is blocked then  a {\vsix horizontal} $F$-profile developes: $\dot M(r) = 0$.

If the central object is magnetized, the magnetic field can disrupt the accretion flow at the radius comparable with the Alfvén radius. 
Thus, the inner radius of disc  depends on the accretion rate.
Moreover,  the inner disc can lack the axial symmetry and one has to take into account the uncertainty of a definition of the inner boundary. 
The instant accretion rate through the disc inner edge is determined by {\vsix a complex of} local MHD, thermal, and dynamical processes. 
The global disc evolution (or $\dot M_{\rm disc} (t)$), however, depends on the inner conditions averaged over a time period comparable to the global viscous time.

The implications for a disc around a neutron star during an outburst are the following. 
Consider a disc with diminishing accretion rate. 
The inner radius is shifting away from a magnetized  star, eventually reaching the corotation radius. 
Suppose that, {\vsix when} $\Rin\gtrsim \Rcor$,  the matter is mostly trapped in the disc near the inner edge and not expelled. 
For example, \cite{dangelo-spruit2010} argue that this would be the case if the inner edge is close to the corotation radius. Then, on the time-scale of the {\em viscous time at the inner radius}, the matter is piling there until it pushes the magnetosphere so that the accretion on the star is possible again. 
The viscous time at the corotation radius in the disc with a semithickness $z_0$ can be estimated as $t_{\vis}(\Rcor) = t_{\rm K}/\alpha/(z_0/r)^2 \sim P_\star\,  (\Rcor/z_0)^{2}/\alpha \sim (10^2- 10^3) P_\star$, where $t_{\rm K} \equiv 2\pi/\omega_{\rm K}$ and $P_\star$ is the star spin period. 
For a millisecond pulsar, $t_{\vis}(\Rcor)$ can be  $\sim 0.1- 1$~s. 
Thus, on such times the matter `brims over' the centrifugal barrier again and again, while there is a flow from the outer disc parts~(a `quasi-propeller' mode as coined by \citealt{Hartman+2011}).  
Analytic descriptions of such oscillations were developed by \citet{Spruit-Taam1993} and \citet{dangelo-spruit2010} in  time-depenedent radial models~(hence the `Spruit-Taam instability'). Furthermore, \citet{Ertan2017,Ertan2018} studied the conditions for propelling the matter from the disc, when its edge is beyond $\Rcor$,
and argued that steady propelling could not happen arbitrary far from  $\Rcor$, thus also bringing about the billowing outflow events.

Oscillating character of the flow stopped beyond the corotation radius was also demonstrated in numerical simulations by
\citet{Romanova+2005,Ustyugova+2006,Romanova+2009,zanni-ferreira2013,Lii+2014,parfrey+2017}, \citetalias{Romanova+2018}.
In particular, \citetalias{Romanova+2018} have shown that the inner disc radius $\Rin$ oscillates around some value:  {\vsix when} $\Rin$ is far enough,  the matter accumulates; the radius gradually decreases and provokes a spike of ejection (and sometimes accretion) associated with a magnetic-lines inflation. 
The  accretion events are highly asymmetric relative to the vertical coordinate $z$ and the azimuth.
\citetalias{Romanova+2018} also report that the inner radius is varying, permitting accretion,  on time scales of 0.1~s when scaled to typical parameters of AXMPs. 
These variations cannot be seen in the outburst long-term light curves, but, {\vsix presumably,  they can be observed} on proper time-scales. 

To sum up, if one sets an effective averaging condition at the inner radius, the whole disc evolution can be calculated by solving Eq.~\eqref{eq.diffusion}.
For this we use the code {\freddi}~\citep{Malanchev-Lipunova2016} developed for modeling  burst light curves of BHXTs, utilizing the
solution for vertical structure  of a geometrically-thin disc in an analytic form (see Appendix~\S\ref{Appendix.OPAL_vertical_structure}).
In further subsections we describe  new features of the code.

\subsection{Inner radius of the disc}\label{ss.Rin_def}

If the accretion rate on a magnetized neutron star is high, the accretion disc can reach the star surface or spiral down to it from the radius of the last stable orbit~\citep[see, e.g.,][]{sunyaev-shakura1986}.
Otherwise,  the disc is truncated by the magnetosphere.

Models of discs  threaded by magnetic lines assume that the inner radius can be found by comparing  the disc-height integrated magnetic stress with the angular momentum flux,
see, e.g.~\citet{Wang1996,armitage-clarke96,matt-pudritz2005,kluzniak-rappaport2007}. 
For diamagnetic discs, a pressure balance is employed instead~\citep[e.g.,][]{elsner-lamb1977,sunyaev-shakura77e,lipunov1978,Aly1980,Chashkina+2017}. 
Both approaches yield the same expression  for $\Rin$, within some  dimensionless factor.
We choose for the present work the following parametrization:
\begin{equation}
\rin= \xinner \, R_\mag\, , \quad {\rm where} \quad
 R_\mag \equiv  \left(  \frac{\mu^2}{\dMout \, \sqrt{G \, \mns}}\right)^{2/7}\,  .
 \label{eq.rmag}
\end{equation}
One can regard  parameter $\xinner$ as the one incorporating our current uncertainty about the physics at the disc-magnetosphere boundary. 
There is also an issue of matching  a  1D model with a real  picture of asymmetric and fluctuating flows of plasma. 

Estimates of $\xinner$ obtained in 1D models  are: 0.5 \citep{ghosh-lamb1979iii}, $0.3-1$~\citep{Chashkina+2017}, $\xinner<1.5$ and it is a function of $\omega_s$~\citep{kluzniak-rappaport2007}. 
MHD simulations  give $\xinner \sim 0.4-0.5$ \citep{Long+2005,Bessolaz+2008} for fixed $\mu$ and $\dot M$, whereas a different dependence on  these parameters, comparing to \eqref{eq.rmag}, is formulated by \citet{kulkarni-romanova2013}.

If the {\vsix inner radius} $\Rin$ exceeds the radius of the light cylinder 
\begin{equation}\label{eq.Rlight}
R_{\rm light}=\frac{c}{\omega_\star}\, ,
\end{equation} 
the quasi-stationary disc pressure cannot balance the pressure of the pulsar wind: the former decreases as $R$ in power of $-2.5 ..-3$ and the latter as  $R^{-2}$~\citep{Shvartsman1970,lipunov1992}; apparently, the disc material is to be dispersed.

\subsection{Accretion rate on the neutron star}\label{s.Mdot_ns}

The accretion rate at the disc inner boundary is determined by the disc evolution which, in turn, depends on the disc mass, viscosity, and boundary conditions. 
In the standard accretion regime, all the matter reaching the disc inner edge is transferred to the star (if $\Rin<\Rcor)$.

If the disc inner radius exceeds a corotation radius
\begin{equation}
\rc = \left( \frac{G \mns P_{\star}^2}{4\pi^2} \right)^{1/3} \approx 1.7 \times 
10^6\, m_{1.4}^{1/3} \, P_{-3}^{2/3}~ \mbox{cm}\, ,
\label{eq.Rcor}
\end{equation}
where $P_{-3} = P_{\star}/0.001$~s, the centrifugal force, acting on the matter rotating with $\omega_\star=2\,\pi/P_\star$ in the magnetosphere, is larger than the gravitational pulling at the equator, thus inhibiting the matter falling to the neutron star. 

One can imagine that a total blockage of accretion onto the neutron star occurs when $\Rin$ becomes equal or larger than $\Rc$. If so, the matter reaching the inner disc radius is either  expelled from the system, or accumulated in the disc, or both. 
The former regime corresponds to a 'propeller'~\citep{Illarionov-Sunyaev1975}, the latter, to a so-called `dead disc'~\citep[][]{Sunyaev-Shakura1977}.
\begin{figure}
    \centering
    \includegraphics[width=0.45\textwidth]{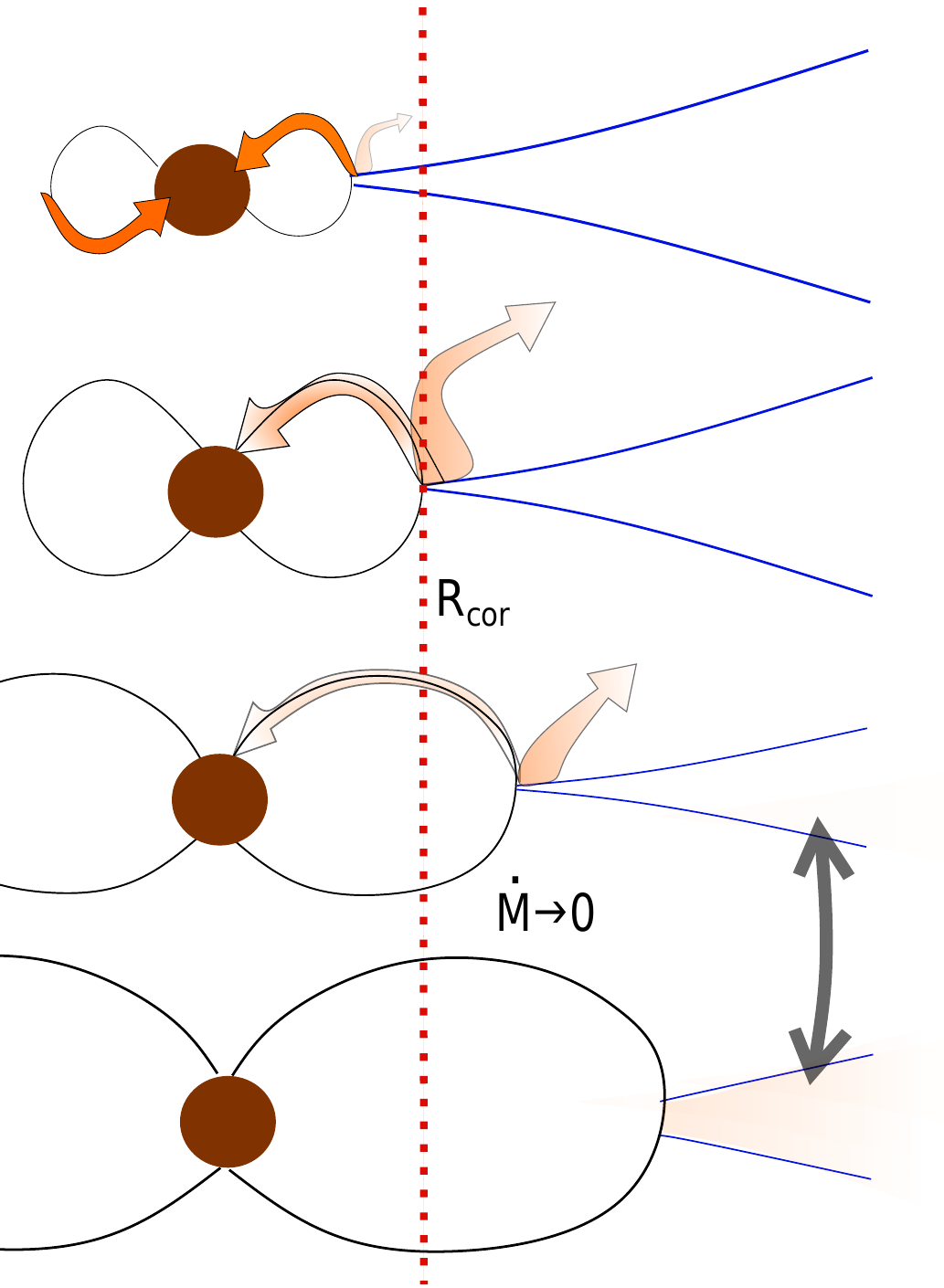}
    \caption{Illustration to the \PONS{} scenario. The two lower pictures illustrate the propeller regime, or rather its two snapshots, assuming   the oscillatory character of the accretion flow. The inner disc  `breathes' quasi-periodicaly between the two  extreme states.  The  \PONS{} scenario implies that some matter can outflow even when the inner disc radius $\Rin<\Rcor$.
    }
    \label{fig:accretion_scenario}
\end{figure}
However, the real geometry and physics of the problem is quite involved and it would be  reasonable, and  in line with various analytical and numerical results, to assume that some matter can reach the neutron star even when  $\Rin>\Rcor$~(see Fig.~\ref{fig:accretion_scenario}).For example, 2D simulations of \citetalias{Romanova+2018} show that a part of the matter can flow above or below the closed part of the magnetosphere and accrete onto the star~\citep[see also][]{zanni-ferreira2013}.  {\vseven \cite{Gungor+2017} have suggested that a fraction of the accreted matter may be determined from
an analysis of the light curves assuming that an underlying disc accretion-rate time-dependence is known. In the present study, we choose another strategy:  to obtain the variation of the disc accretion rate making most simple assumptions about developments at the inner boundary.
}

\begin{table*}
\caption{Reference to model scenarios of accretion onto neutron star, see \S\ref{s.Mdot_ns} and  \S\ref{s.Fin}. }    \label{tab:model_scenarios}
\begin{tabular}{l|c|c|c|c}
Scenario &  Concept &   Accreted part &  \multicolumn{2}{c}{Torque coefficient}  \\
& &$f_\star = \dot M_\star/\dMin$ & for $\Rin < \Rcor$   & for $\Rin > \Rcor$ \\  
 \hline
 \NS & no-outflow, everything falls onto NS & 1  &$\kappat=\xinner^{7/2}$ &$\kappatd = \kappat$  \\
 \PO & propeller outflow; nothing falls onto NS if $\Rin>\Rc$ &  $H(\Rc - R_\mathrm{in})$& $\kappat=\xinner^{7/2}$ & $\kappatd = 0$       \\
 \PONS & gradual blocking, matter partly outflows   & 
 $ f_\star(\omega_s)$ &$\kappat=\xinner^{7/2}$ &
 $\kappatd = \kappat$ \\
 \hline
 \end{tabular}
\end{table*}
Altogether, the matter flow can be divided in three ways: down to the neutron star, away by the propeller-mechanism, back to the disc. 
In our code, we introduce a penetration parameter 
$f_\star  = \dot M_\star/\dMin$, which can be set according to different schemes\footnote{SA -- 'star accretion'; PO -- 'propeller outflow'; POSA -- both} of matter penetration~(see Table~\ref{tab:model_scenarios}):
\begin{itemize}[leftmargin=0.4in]
    \item[    `\NS'] There is no outflow at the magnetospheric boundary. All  matter is either accreted on the  star in the course of short-term events  or remains in the  disc.
    If $\dMin$ is not zero then it is has to fall on the neutron star.

    \item[`\PO'] Total screening of the neutron star by the magnetosphere if its radius exceeds the corotation radius. Then the penetration parameter is the  step Heaviside function \hbox{$f_\star =  H(\Rc - R_\mathrm{in})$}.
    
    \item[`\PONS'] Gradual blocking of accretion onto the neutron star.  In the present work, we use an approximation to numerical simulations by \citetalias{Romanova+2018} that provides  $f_\star$ as a function of the fastness parameter $\omega_s \equiv \omega_\star/\omega(\Rin)$. 
 \end{itemize}
\citetalias{Romanova+2018} numerically found a {\vsix time-averaged} efficiency of a propeller, which depended on the poloidal wind velocity. 
If the accretion flow at the disc edge  is split into the accretion flow onto the star and the wind, then the portion of the accretion rate onto the star is  \hbox{$f_\star =  1-f_{\rm eff}$}, where $f_{\rm eff}$ is a parameter of \citetalias{Romanova+2018}, which depends on the characteristic velocity of the outflowing matter regarded as the wind.
If the wind  contains the matter with the poloidal velocity {\vsix higher} than $v_{\rm min}=v_{\rm esc}$, where the escape velocity $v_{\rm esc}=\sqrt{2G\mns/r}$, then  $ f_{\rm eff}=0.0006\,\omega_s^{4.01}$~(see table 2 of \citetalias{Romanova+2018})\footnote{ Precision of the power index is excessive in our context but we keep it identical with that in \citetalias{Romanova+2018}.}.
The portion of the  propelled matter and the average velocity of the outflow become larger  with increasing fastness $\omega_s$. 
The fate of the lower-speed outflow is obscure; it may flow back to the disc\footnote{See also a discussion on failed magnetic expulsion and circulation in the disc by~\citep{Spruit-Taam1993}.}, but that was beyond the simulation region of \citetalias{Romanova+2018}. {\vseven Presumably, the matter outflowing from the inner boundary with a mild velocity returns back to the disc and reaches the inner boundary again after the viscous time at the radius of return. Generally, this time scale is expected to be shorter than the
viscous time scale of the whole viscously-evolving  disc, and such circulations are effectively averaged out in the long-time
picture\footnote{In the majority of specific models, described in \S\ref{s.results}, $R_{\rm hot}$ is at least two orders of magnitude larger than $\Rin$ when $\Rin<R_{\rm light}$.}. However, if the hot zone had small radial extent, such circulations would affect its dynamic viscous evolution.}


Each of the above scenarios should be regarded as a toy model, allowing us to make conclusions about the overall disc evolution.

\subsection{Torque at the inner edge of a  disc}\label{s.Fin}

Outside the corotation radius, the magnetic torque accelerates the rotation of the matter or pushes the disc matter outward, depending on the diamagnetic properties of the disc. 
In either case, assuming that interaction zone is not too wide, the inner boundary condition on the viscous torque can be written as:
\begin{equation}\label{eq.Fin_2part}
\Fin =  \kappatd \,\frac{\mu^2}{\rin^3}  \mbox{~~for~~}\Rin>\Rcor\,,
\end{equation}
\citep{davidson-ostriker1973,lyn-pri1974,sunyaev-shakura77e,lipunov1992,Spruit-Taam1993,kluzniak-rappaport2007,matt-pudritz2005}. 
Here $\kappatd$ is the dimensionless coefficient incorporating  our uncertainty of MHD processes at the boundary between the disc and magnetosphere.

{\vsix Inside the corotation radius, the magnetic torque {\vseven is expected} to decelerate  rotation of the disc, `helping' the  viscosity to remove angular momentum from the matter, so that the accretion process is facilitated. 
Consequently, the release of the viscous heat in the disc is to be modified.
The size of the region, where the disc and  magnetic field effectively interact, affect the actual values of the total magnetic torque and the location of the inner disc edge \citep[for example,][]{ghosh-lamb1979iii,wang1987, wang1995,armitage-clarke96,matt-pudritz2005,kluzniak-rappaport2007}.   
As an approximation for a disc with a relatively thin interaction zone, the same view of $\Fin$ as \eqref{eq.Fin_2part} can be adopted for the accretion regime, but with a fixed radius:
\begin{equation}\label{eq.Fin_1part}
\Fin =  \kappat\, \frac{\mu^2}{\rc^3}  \mbox{~~for~~} \Rin<\Rcor\, .
\end{equation}
This condition numerically approximates the situation of the viscous torque approaching zero, $\Fin\rightarrow 0$, since $\Fin/(\dMin\hin) \propto \kappat (\Rin/\Rcor)^3 \rightarrow 0$ for $\Rin\ll\Rcor$. 
Alternatively, as a next approximation, one could solve Eq.~\eqref{eq.diffusion} with a radially-distributed magnetic torque to resolve the structure of the interaction zone (see the previous references).

Notice that, while the torque coefficient can be set different in \eqref{eq.Fin_2part} and \eqref{eq.Fin_1part}, in the  modeling below we keep $\kappatd=\kappat$ in {\vsix both} \NS and \PONS scenarios to ensure the smoothness of a solution.}

If the matter is propelled away from $\Rin$, the work done by the magnetic torque  can be `divided' between the propelled matter and the matter remaining in the disc.
Various models suggest that propelling is hardly very effective when $\Rin$ is not much greater than $\Rc$, and thus, generally, $\kappatd$ is a function of the fastness  \mbox{$\omega_s$~\citep[e.g.,][\citetalias{Romanova+2018}]{dangelo-spruit2012}}. 
For a `very efficient propeller' regime (\PO scenario) we set a simple condition of a zero disc torque: $\kappatd=0$.

\begin{figure}
    \centering
    \includegraphics[width=0.45\textwidth]{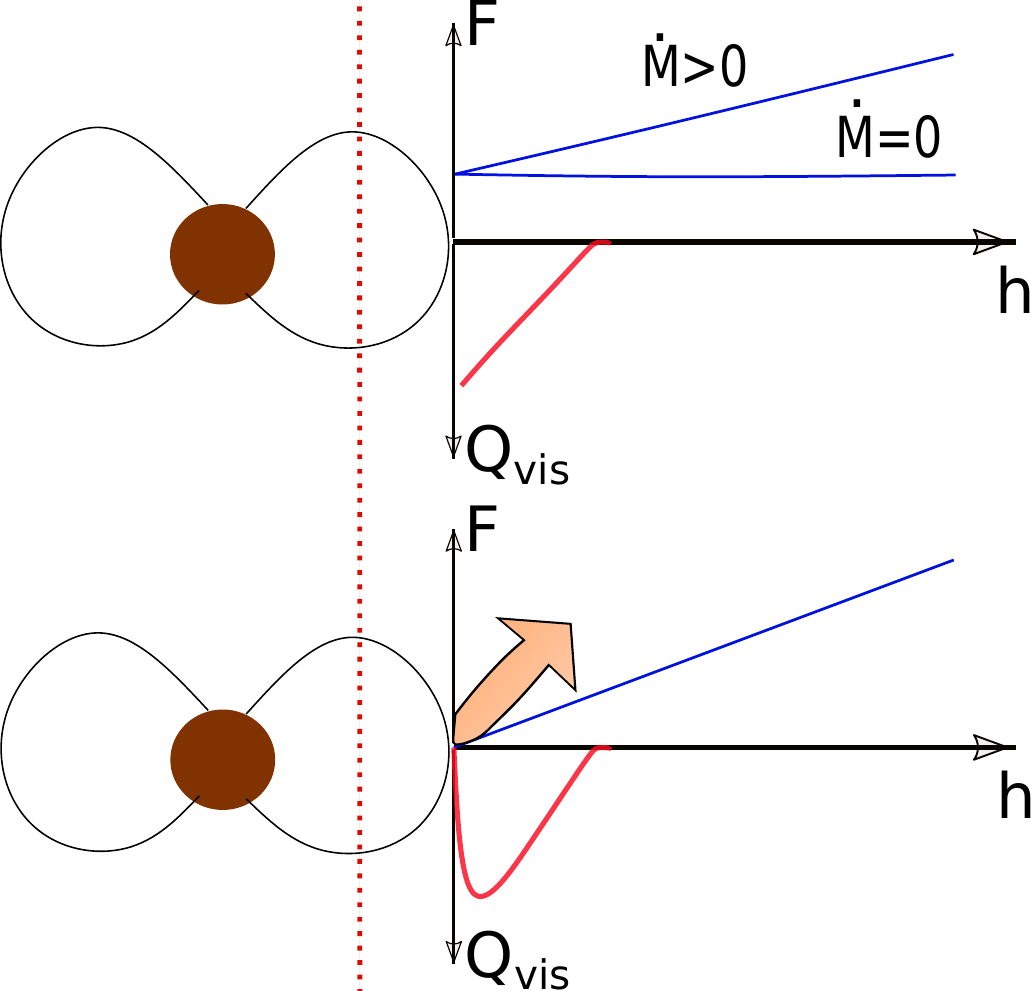}
    \caption{Possible torque distributions $F(h)$ if $\Rin>\Rcor$ are shown by the blue lines. In the top plot, the case with a finite inner torque, corresponding to  \NS and \PONS scenarios, is shown.  In the lower plot, a torque distribution for the \PO scenario with $\kappatd=0$ is shown. Red lines illustrate the heat radiated locally  as the log-scaled distribution $\log(Q_{\rm vis})$ vs. $\log(h)$.
    }
    \label{fig:F_Qvis_h}
\end{figure}
\begin{figure}
    \centering
    \includegraphics[width=0.44\textwidth]{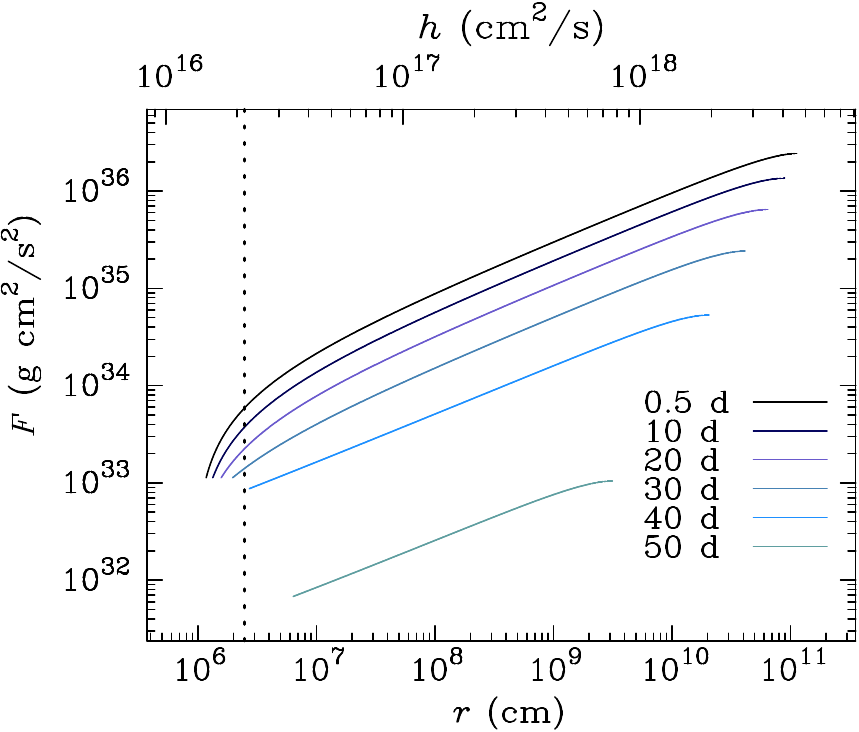}
    \caption{Torque distribution  at different times marked from top to bottom.  Specific angular momentum is shown at the top axis. As the luminosity and  accretion rate drop, the inner radius of the disc moves away, pushed by the magnetosphere. At the same time, the outer radius of the hot zone decreases (see  \S\ref{s.Rout}).
   Corresponding light curves are presented in Fig.~\ref{fig.main_fit}, and the surface density, in Fig.~\ref{fig.disk_Sigma}. The vertical line marks the corotation radius.
    }
    \label{fig:freddi_F_h}
\end{figure}

In Fig.~\ref{fig:F_Qvis_h} we schematically illustrate the possible torque distributions when the inner disc radius is beyond the corotation radius.
In a stationary disc without mass sinks, the  radial distribution of the viscous torque is 
\begin{equation}
F =\Fin + \dot M \, (h-h_\mathrm{in})\, .
\label{eq:F_stat}    
\end{equation}
The heat released in the disc can be found as $Q_{\rm vis}= (3/8\pi)\, (G\,M_\star)^4\, F/h^7$.
Figure~\ref{fig:freddi_F_h} shows  calculated distributions of the viscous torque in a hot-zone model with  varying  inner and outer radii {\vsix for $\kappat=\kappatd$.} Notice that the accretion rate $\dot M = \partial{F}/\partial h$ is quasi-stationary in the inner zone  (notice the log scale) and decreases with time.

{\vsix In line with our simplified model, the parameter $\kappat$ can be related to the ratio $\xinner=\Rin/R_\mag$ to satisfy a specific condition of a neutron star spin equilibrium.}
An  effective formulation for the evolution of neutron star angular momentum   associates the spin-up of the star with the torque added by the accreted matter, and the spin-down, with the  magnetic braking~\citep{lipunov1981a,lipunov1982a,Lipunov1982diagram,lipunov1992}:
\begin{equation}\label{eq.NStorque}
 \frac{ {\rm d} I \omega_\star}{{\rm d} t} = \dot M_{\rm in} \sqrt{G\, \mns\,R_\mathrm{in}}
 - \kappat\, \frac{\mu^2}{R_{\rm t}^3}\, , \qquad 
R_t = \max(\Rin,\Rcor)\, ,
\end{equation}
where $I$ is the moment of inertia of the neutron star, $\omega_\star$ is the neutron star spin velocity, $\dMin$ is the  accretion rate at the inner boundary of the accretion disc.
{\vsix Equation~\eqref{eq.NStorque}  can also be formally applied  when  $\Rin\ll\Rc$ and there is no braking of a neutron star: }
since the 1st term on the r.h.s of \eqref{eq.NStorque} exceeds greatly the 2nd term, the latter  can remain in the sum as a non-important term.

If  a neutron star is in the equilibrium, the l.h.s. of \eqref{eq.NStorque} is zero: the accreted angular momentum, spinnig-up the star, is balanced by the negative magnetic torque, which concurrently spins down the star. 
According to various models, the equilibrium occurs when the inner disc radius is close but slightly less than the corotation radius~\citep{wang1995,armitage-clarke96, kluzniak-rappaport2007, matt-pudritz2005}. 
Substituting the magnetic torque \eqref{eq.Fin_1part} as the second term in r.h.s. of \eqref{eq.NStorque}, one obtains:
\begin{equation}\label{eq.NStorque1}
\frac{ {\rm d} I \omega_\star}{{\rm d} t} = \dot M_{\rm in} \sqrt{G\, \mns\,R_\mathrm{in}}\, \Big(1-\kappat\,\xinner^{-7/2} \Big(\frac{\Rin}{\Rc}\Big)^3\Big)
\end{equation}
Thus for a neutron star to be in the equilibrium when $\Rin\approx \Rc$, the parameters should be related as follows: $\kappat \approx \xinner^{7/2}$. For example, if we set $\xinner =0.5$ then $\kappat \approx 0.088$.

Further, as an option, we shall study a case of a `left-over'  disc.
For this, we shall fix some final value of the magnetospheric radius $\Rd$.
The inner disc radius `freezes' when it expands to the value $\Rd$ and 
\begin{equation}\label{eq.Fdead}
\Fd = \kappatd \,\frac{ \mu^2}{\Rd^3}\, .
\end{equation}
Such `dead' disc, having Keplerian rotation at each radius, radiates the viscous heat,  the energy for which is supplied by the neutron star rotation.

\subsection{Outer radius of the hot disc and irradiation of the disc}\label{s.Rout}
The disc consists of two parts: the `hot' ionized and the `cold' neutral  outer part. 
The irradiation of the disc affects the radius of its ionized  hot  part if the irradiating flux exceeds the local radiation flux due to the viscous heating:
\begin{equation}\label{eq.Cirr}
Q_\irr  = \mathcal{C}_\irr \, \frac{L_x}{4\,\pi\,r^2} > Q_\vis\, ,
\end{equation}
where $\mathcal{C}_\irr$ is a dimensionless irradiation parameter. Under the assumption that $\mathcal{C}_\irr$ is constant and $L_x = \eta_{\rm accr} \dot M c^2$, the ratio of these fluxes depends only on the radius:
\begin{equation}\label{eq.irr2vis}
\frac{Q_\irr}{Q_\vis} = \frac 43\, \eta_{\rm accr} \, \mathcal{C}_{\irr} \, \frac{r}{\Rg}\, ,
\end{equation}
where $\Rg= 2GM_\star/c^2$~\citep{suleimanov_et2007e}.
The `irradiation temperature' can be defined as $Q_\irr \equiv \sigma\, T_\irr^4$.
Following \citet{tuchman_et1990,dubus_et1999}, we adopt the following condition to define  the hot disc outer radius $\Rhot$: 
$T_{\irr}=10^4$~K if \eqref{eq.Cirr} fulfills.

When condition \eqref{eq.Cirr} deteriorates at $\Rhot(t)$ at some late time, the irradiation-controlled evolution is superceded by an evolution determined by the cooling-front propagation, in accordance with the DIM model~(\citealt{dubus_et2001,lasota2001,Hameury2020}; see also \citealt{lipunova-malanchev2017}).  
To find $\Rhot(t)$ one can use the cooling front velocity. 
The cooling front velocity evolves in a complex fashion, modeled in the dedicated studies \citep[e.g.,][]{menou_etal1999fronts}.  
We adopt an analytic approximation to the cooling front velocities found by \citet{Ludwig+1994}, see their section 3. 
In such a framework, the front velocity depends on the values of $\alpha_{\rm hot}$ and $\alpha_{\rm cold}$, mass of the central star, current radius, and the local column density.  
Consequently, we obtain a significant acceleration of the hot-disc boundary propagation when $Q_\irr$ becomes less than $Q_\vis$ at $\Rhot$.

Thus, irradiation parameter has a great effect on the disc long-term dynamical evolution.
The following analytic expression~\citep[e.g.,][]{suleimanov_et2007e}  
\begin{equation}\label{eq.Cirr_theor}
\mathcal{C}_\irr = (1-A)\,  \Psi(\theta) \, \frac{z_0}{R} \, q\, ,{\rm ~~where~~} q=\left (\frac{{\rm d}\ln z_0}{{\rm d}\ln r} -1 \right)
\end{equation}
 could be used to estimate  $\mathcal{C}_\irr$ in \eqref{eq.Cirr}, where $(1-A)$ is the portion of absorbed and thermally reprocessed incident flux, $z_0$ is the semithickness of the disc, and $\Psi(\theta)$ is the angular distribution of the irradiating flux, where $\theta$ is the angle between a ray from the centre and the normal to the disc.
 For the gas-pressure dominated, absorption-opacity zone 
 {\vsix (with the Kramers law)} 
of the disc, $q=1/8$~\citep[e.g.,][]{shakura-sunyaev1973}.

Unfortunately, a problem with estimate  \eqref{eq.Cirr_theor} is that resulting $\mathcal{C}_\irr$  is too low, if one considers direct irradiation of the photosphere of a standard disc.
For example, \citep{suleimanov_et1999} estimated $(1-A) \sim 0.1$. 
Substituting $z_0/r =0.06$ into \eqref{eq.Cirr_theor}, we get $\mathcal{C}_\irr \approx 7.5\times 10^{-4} $.
On the other hand, it is well established~\citep[e.g.,][]{suleimanov_etal2008} that $\mathcal{C}_\irr$ must be  larger to explain optical flux of LMXBs. 
A hypothesis was proposed that  an extra scattering occurs  in the medium above the disc and enhances the irradiation factor. 
\citet{mescheryakov_etal2011} considered the radiation transfer problem for a disc with an extended atmospheric layer around a neutron star. 
They assumed that the  disc was irradiated  by the central flux from the neutron star with $L = \eta_\star \, \dot M \,c^2$ with the accretion efficiency  $\eta_\star =0.1$ and  angular distribution \hbox{$\Psi(\theta)=1$.}
Following the graphical results of their figure 8 and taking into account \eqref{eq.irr2vis}, we deduce that the characteristic value of $\mathcal{C}_\irr \approx (4-8)\times 10^{-3}$ for $L\approx \Ledd$ and $\mathcal{C}_\irr\approx (2.6-3.5)\times 10^{-3}$ for $L\approx 0.1\, \Ledd$ at $R\sim 10^{11}$~cm.

One can use the optical data to  constrain $\mathcal{C}_\irr$. 
This is what we do for Aql\, X-1 in the course of data fitting.
We take into account  the fact that the angular distribution of the X-ray flux from the star differs from that of the flux coming from the central disc. 
We assume that the star radiates isotropically, $\Psi(\theta)=1$.  For radiation from the disc, we take
$\Psi(\theta)=2\, \cos(\theta) \approx 2\, z_0/r$ ~\citep[e.g.,][]{suleimanov_et2007e}, assuming that {\vsix the central part of the disc lie in the equatorial plane of the outer disc}.

In the model we set two different parameters $\widetilde{\mathcal{C}}_\irr$, for the cold and hot part of the disc, which relate to parameter $\mathcal{C}_\irr$ from \eqref{eq.Cirr} as follows
\begin{equation}
\begin{split}
&\mathcal{C}_\irr = \widetilde{\mathcal{C}}_\irr \, \left(\frac{z_0/r}{0.05}\right)^k \Psi(\theta), \qquad k=1\, ; \\
 &\mathcal{C}_\irr^{\rm cold} = \widetilde{\mathcal{C}}^{\rm cold}_\irr  \,  \Psi(\theta)\, .
 \end{split}
\label{eq.Cirrboth}
\end{equation}

Both disc parts can contribute to the optical flux. 
However, only the `hot' part of the disc (with $T_\eff > 10^4$~K) is involved in the fast viscous evolution; moreover, its size determines the global viscous time, i.e., the rate of the evolution.
The viscous evolution in the colder part proceeds on a much longer viscous time scale because the temperature  and viscosity are lower there. 
Additionally, slower evolution may be explained  by suggested lower values of $\alpha$ in the disc with recombined material \citep[e.g.,][]{smak1984a}.
{\vfour At the outer boundary of the hot part we assume $\dot M = 0$~\citep{lipunova-malanchev2017}. Comparing to models with a resolved  cooling front structure~\citep[e.g.,][]{dubus_et2001}, our simple approach may result in a skewed estimate of the parameter $\alpha$.}

\subsection{Observed flux from  the disc and neutron star}\label{s.flux_disc_NS}
Thermalization of the incident flux in the disc upper layer leads to the enhancement of the local black-body radiation. Subsequently, the local effective temperature can be expressed as 
\begin{equation}\label{eq.Teff}
T_{\eff}^4 = T_\irr^4 + T_\vis^4\, .
\end{equation}
Observed spectral flux is calculated taking into account possible color correction:
\begin{equation}
  \mathcal{F}_{\nu,\mathrm{disc}} = 
  \int I_\nu \, {\rm d}\Omega = 
  \frac{2\,\pi \cos{i}}{d^2\,f_{\rm col}^4 } 
    \int_{R_\mathrm{in}}^{R_\mathrm{out}} r\, B_\nu(f_{\rm col}\,{T_\mathrm{eff}(r)})\,\mathrm{d}r\,,
\label{eq.obs_flux_at_nu}    
\end{equation}
where {\vseven ${\rm d}\Omega = 2\pi r{\rm d}r \cos{i}/d^2$ is the solid angle, at which an observer sees a disc ring,}  $i$ is the disc inclination to the line of sight, $d$ is the  distance to the source, $T_\mathrm{eff}$ is obtained according to \eqref{eq.Teff}, and {\vseven $I_\nu$ is the ring intensity.}
In the photosphere of the disc electron scattering can modify the blackbody spectrum~\citep{shakura-sunyaev1973,Taam-Meszaros1987}. The color correction factor $f_{\rm col}$ 
approximate effects of electron scattering in the disc photosphere~\cite{shi-tak1995,davis_et2005}:  $I_\nu = B_\nu (f_{\rm col} \,T_\mathrm{eff})/f_{\rm col}^4. $
Color correction factor $f_{\rm col}$ is set to unity for the optical and IR bands and to 1.7, for X-ray.
We ignore effects of irradiation when calculating flux in an X-ray band.

The viscous heat flux determines $T_\vis$ as follows:
\begin{equation}\label{eq.Tvis}
\sigma\, T_\vis ^4 = \frac{3 \, (G\,M_\star)^4\, F}{8\,\pi\, h^7} \, ,
\end{equation}
where the torque $F$ is taken as the solution to the viscous evolution equation \eqref{eq.diffusion}.
In the central parts of the disc, the heat released by the viscosity  dominates largely any irradiation.

{\vseven For the hot part of the disc, we calculate \eqref{eq.obs_flux_at_nu}     between $\Rin$ and $\Rhot$, and for the cold part, between  $\Rhot$ and the tidal radius $R_{\rm tid}$, where $R_{\rm tid}$ equals 90\% of the Roche lobe radius, close to the the result by \citet{pap-pri1977}.}
For the cold part of the disc we take \hbox{$T_{\rm vis}=0$} in \eqref{eq.Teff} since {\vdva we are not interested in the solution for the vertical structure of the  cold disc.} 
{\vsix Its relative semithickness is a parameter of \freddi and is set to 0.05 in the current work.}
All connected uncertainty is believed to be parametrized by $\widetilde{\mathcal{C}}_\irr^{\rm cold}$. 
    
{\vsix The bolometric luminosity of a viscously-heated disc can be found by integrating the radiative flux \eqref{eq.Tvis} over the disc surface. {\vsix Since the main energy release takes place  close to the centre, we can substitute the outer radius by infinity. Notice also that close to the centre  the stationary solution \eqref{eq:F_stat} holds very accurately. One arrives at:}
\eqref{eq:F_stat}:}
\begin{equation}
L_{\rm disc} =\Big(F_\mathrm{in} + \frac{\dot M_\mathrm{in} \,h_\mathrm{in}}{2}\Big)\, \times \,  \omega_\mathrm{in} \, 
\label{eq.L_bolometric}    
\end{equation}
in the Newtonian case, where $\omega_{\rm in}$ is the Keplerian frequency at the inner edge of the disc, and $h_\mathrm{in} = \omega_{\rm in} \, \Rin^2$.  Subsequently, disc accretion efficiency is 
\begin{equation}\label{eq.eta_disc}
\eta_{\rm disc} = \frac{F_\mathrm{in}\,\omega_{\rm in}}{\dot M_\mathrm{in}\,c^2} + \frac {G\mns}{2 \Rin c^2}\, .
\end{equation}

In the Newtonian mechanics, the accretion efficiency of the neutron star $\eta_\star \equiv L_\star/\dot M_\star/c^2$ is found {\vsix assuming that  the kinetic energy of radial motion of matter at the star surface is converted to radiation}:
\begin{equation}
\eta_\star = \frac{\Rg}{2\,R_\star} \, \left(1 - \frac{R_\star}{\Rin}\right)
+   \frac {\Rg}{4\,\Rin} \, \left[ 1- 2\, \omega_s + \omega_s ^2
\left(\frac{R_\star}{\Rin}\right)^2\right]\, ,
\label{eq.general_eta_ns}   
\end{equation}  
where $\omega_s \equiv \omega_\star/\omega_{\rm K}(\Rin)$~(see Appendix \ref{s.bolLns}).

We assume that the emitting region occupies some part of the neutron star surface, $S_\mathrm{X} / (4\pi \, R_\star^2)$, and the spectrum is black-body.
The effective temperature of this region is $T_\star = [L_\star / (\sigma_\mathrm{SB}\,S_\mathrm{X})]^{1/4}$.
If the radiation is isotropical, the observed {\vtri  flux density is}
\begin{equation}
  \mathcal{F}_{\nu,\star} = \frac{S_\mathrm{X}}{4\pi \, d^2} 
    \pi B_\nu({T_\star})\,.
\end{equation}

Since we are interested in burst light curves, that is, in the flux variations over time of days, the possible oscillations on {\vsix time} scales $\ll 10^4-10^5$~s are smoothed  out. In other words, the accretion rate in \eqref{eq.L_bolometric} and \eqref{eq.Lstar} is an `average' value determined by the viscous evolution of the disc, i.e., by the solution to equation \eqref{eq.diffusion}.

\subsection{Flux from the companion star} \label{S.flux_Mopt}
\begin{figure}
    \centering{
    \includegraphics[width=0.5\textwidth]{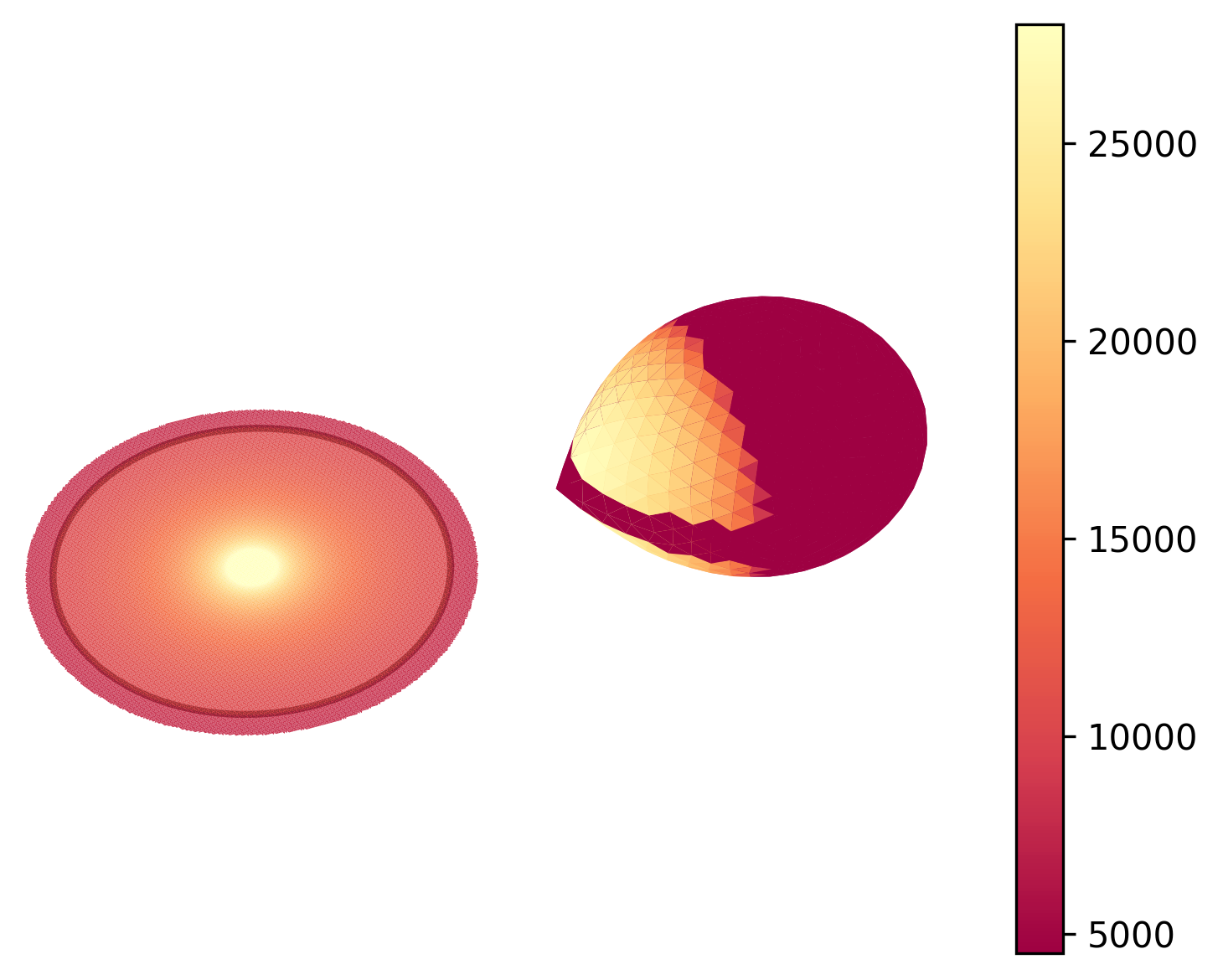}}
    \caption{Visualization of the binary system {\vsix with the parameters listed in Table~\ref{tab:AqlX1_pars} and accretion rate equal to the peak one in model \SAmain (see Table~\ref{tab:result_parametersshort}).} The circle on the disc shows the border between hot and cold zones.
    Effective temperature is color-coded. }
    \label{fig:3dsketch}
\end{figure}
Our code calculates  orbit-modulated light curves of the companion star, irradiated by  X-rays. Fig.~\ref{fig:3dsketch} presents a visualization of the  resulted effective temperature on the disc and the optical star in Aql\,X-1.
The effective temperature of the disc is calculated according to \eqref{eq.Teff}, and 
\begin{equation}\label{eq.Teff_star}
T_{\eff}^4 = T_\irr^4 + T_{\rm opt}^4\, .
\end{equation}
for the star, where $T_{\rm opt}$ is the effective temperature of a non-irradiated star, and
\begin{equation}\label{eq.Tirr_star}
\sigma\,T_\irr^4 =  \frac{(1-a_{\rm opt})( L_{\rm disc} \, \Psi(\theta) + L_{\star})}{4\,\pi\,r^2}\cos{\xi}\, ,
\end{equation}
where $a_{\rm opt}$ is the star's bolometric albedo, $r$ is the distance between the neutron star and a point on the surface of the optical star, and $\xi$ is the angle between the direction to the neutron star and normal to the optical star surface. 
Other designations are as in \S\ref{s.Rout}.
The star surface is calculated using the Roche lobe filling factor $R_{\rm pol} / R_{\rm pol}^{\rm Roche}$~\citep{Antokhina1988}. Resulting orbit-modulated light curves were compared with those obtained by a more sophisticated code, developed for a twisted disc in~\citet{kolesnikov+2020}, and demonstrated an agreement to an accuracy of 5\%.

\section{Multiwavelength data of the Aql~X-1 outburst in 2013}\label{s.data}

Table~\ref{tab:AqlX1_pars} summarizes parameters which we use for modeling of the 2013 outburst in Aql\, X-1. 
When references are given, we adopt the values from previous modelings or the values lying in the observational limits.
For example, \citet{matasanchez+2017} using their near-infrared observations refined $T0$
and the orbital period  $P_{\rm orb}$
and gives estimates of distance $d = 6\pm2$~kpc, mass ratio $0.41\pm0.08$, inclination $36^{\rm o} < i < 47^{\rm o}$ and spectral class of the donor K4$\pm$2. {\vsix We assume $d = 5$~kpc following \citet{meshcheryakov18}.}

Figures~\ref{fig:flux_evolution_ergs} and ~\ref{fig:flux_evolution_counts} show the observed evolution  during the outburst of Aql\,X-1 in 2013 observed by Swift/XRT, Swift/UVOT, and SMARTS/IR. 
A description of Swift/XRT light curves production can be found in \citet{Evans+2007}.
We performed spectral fits to Swift/XRT data in 0.5-10~keV using a  model  of a black-body plus a power-law tail (see \S\ref{ss.XRT} and Fig.~\ref{fig:sppar_evolution}). 
In the optical, we made magnitude--flux transformation, using passbands zero points given in Table~4 of \citet{meshcheryakov18}~(see \S\ref{ss.optical}).

\subsection{Swift X-ray data}\label{ss.XRT}

\begin{table}
\caption{Fixed parameters of the model of Aql\,X-1.}
     \hskip -0.6cm
    \begin{tabular}{llll}
         Variable & Parameter & Value & Ref.\\
         \hline
         $\mns$ &  NS mass & 1.4~$M_\odot$ &\\
         $R_\star$ &  NS radius & $1.12 \times 10^6$~cm&\\
         $\nu_\star$ &  NS spin frequency & 550 Hz & (1) \\
         $\Rcor$ & Corotation radius & $2.5 \times 10^6$~cm& Eq.~\eqref{eq.Rcor}\\
         $R_{\rm light}$ & Light cylinder radius & $8.7 \times 10^6$~cm&Eq.~\eqref{eq.Rlight}\\
         $R_{\rm tid}$ & Tidal radius of the disc & $0.9 R_{L1}=1.87 R_\odot$ & (2)\\
         $P_{\rm orb}$ & Orbital period & $0.7895126$ & (1) \\
         $T0$ & Ephemeris & 2455810.387~d & (3)\\
         $q$ & Mass ratio  & $0.39$ & (3,4) \\
         $a$      & Semi-axis & 4.5~$R_\odot$& \\
         $M_{\rm opt}$ &  Optical star mass & 0.55~$M_\odot$& \\
         $T_{\rm opt}$ & Optical star temperature & 4500~K & (3) \\
         $a_{\rm opt}$ & Optical star albedo & 0.5 & (5)\\
         $R_{\rm pol} / R_{\rm pol}^{\rm Roche}$ & Roche lobe filling & 1 & \\
         $d$ & Distance to the source & 5~kpc & (3,4,6)\\
         $i$ & Inclination of the orbit & $40^{\rm o}$ & (3) \\
         $E(B-V)$ & Color excess & $0.64\pm 0.04$ & App.\ref{sss.redening}\\
         $\xinner$ & $\Rin/\Rm$ & 0.5 & Eq.~\eqref{eq.rmag}\\
         $\kappat$ & Magnetic torque coef. & $\xinner^{7/2}\approx 0.088$& \S\ref{s.Fin}\\
         $f_{\rm col}$ & Inner disc color correction & 1.7 &\S\ref{s.flux_disc_NS}\\
         $S_\mathrm{X} / (4\pi \, R_\star^2)$ & Hot spot fraction & 0.4 &\S\ref{s.flux_disc_NS} \\
         $(z/r)_{\rm cold}$ & {\vsix Outer disc relative}  & 0.05 & \\
          & semithickness &  & \\
         \hline
    \end{tabular}
        (1) \citet{white-zhang1997,zhang+1998,casella+2008};  (2) \citet{pap-pri1977}; (3) \citet{matasanchez+2017}; (4) \citet{meshcheryakov18}; (5) \citet{Basko+1974};  (6) \citet{Galloway+2008}
    \label{tab:AqlX1_pars}
\end{table}
\begin{figure}
    \centering
    \includegraphics[width=0.45\textwidth]{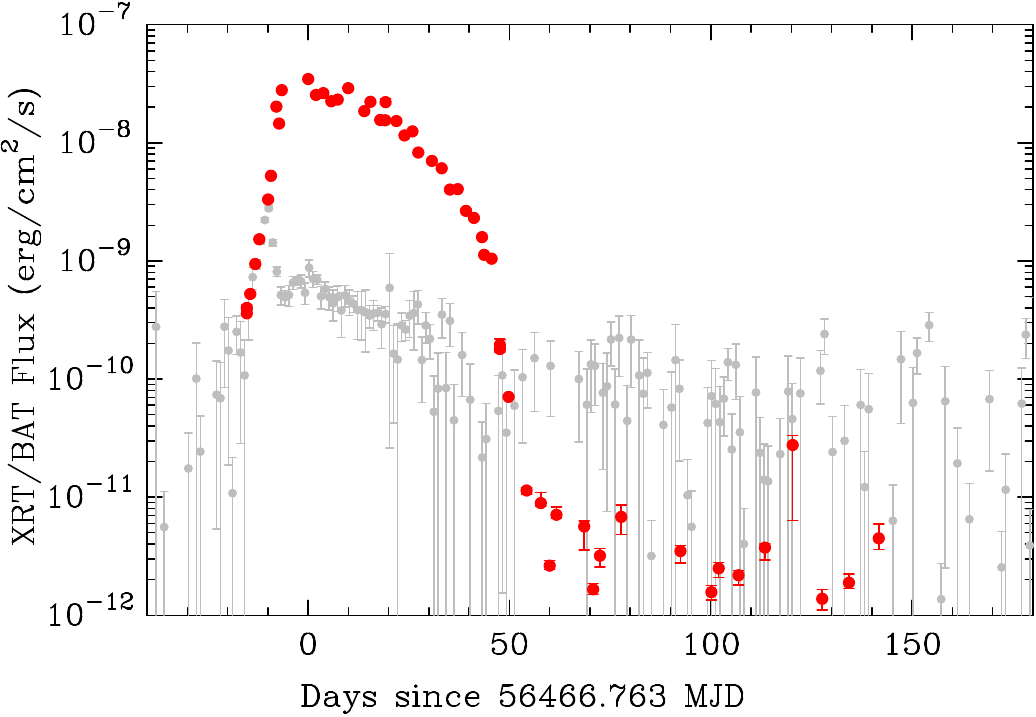}
    \caption{Red circles show unabsorbed total flux in $0.5-10$~keV, observed by Swift/XRT in `wt' and `pc' mode. The flux is calculated using spectral fitting by  model {\tt tbabs*(bbody+powerlaw)} in XSPEC12.10.0c (see \S\ref{ss.XRT}).
     Grey data points show absorbed 15-50~keV Swift/BAT counts converted to absolute units using Crab (see \S\ref{ss.XRT}). 
    }
    \label{fig:flux_evolution_ergs}
\end{figure}
\begin{figure*}
    \centering
        \includegraphics[width=0.86\textwidth]
    {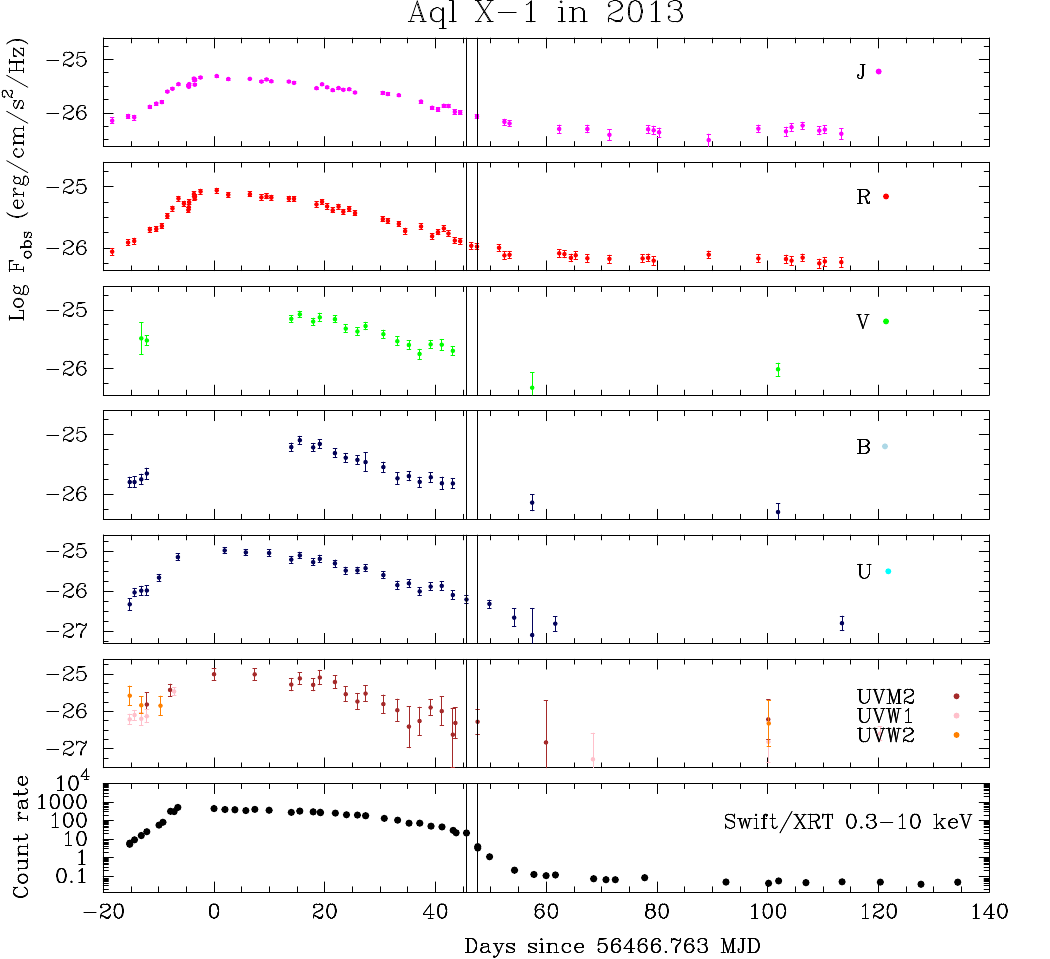}
    \caption{{\vdva From top to bottom: logarithm of deredened {\vtri flux density}, SMARTS/IR ($J$ and $R$) and Swift/UVOT ($V$, $B$, $U$, and ultraviolet bands),  the Swift/XRT 0.3$-$10~keV light curve  as counts/sec, `wt' and `pc' mode. The zero time is set to MJD~56466.763 (23 Jun 2013). Two vertical lines mark the `knee position' -- the time interval when the X-ray flux starts to drop rapidly.}}
    \label{fig:flux_evolution_counts}
\end{figure*}

Spectra, produced by the automatic pipeline in $0.3-10$~keV~\citep{Evans+2009}, as well background, and ancilliary response files were obtained from the UK Swift Science Data Centre using the Swift-XRT data products generator\footnote{\url{www.swift.ac.uk/user_objects}}.
These spectra were rebinned so they had minimum of 1 count in each bin.
Subsequently, we have performed spectral modelling by black-body and power-law components in XSPEC12.10.0c for observations in `wt' and `pc' mode for 0.5-10~keV using the C-statics. For the absorption we set {\tt abund wilm}, and  photoionization cross-sections {\tt xsect vern}.
The unabsorbed flux in 0.5-10~keV is calculated using  model {\tt tbabs*cflux(bbody+powerlaw)}. Uncertainty ranges (1-$\sigma$) on the intrinsic flux are obtained by the Marcov-Chain Monte-Carlo simulations.

Red circles in Fig.~\ref{fig:flux_evolution_ergs} show resulting evolution of the total intrinsic flux when all spectral parameters are free to vary. 
Fig.~\ref{fig:sppar_evolution} shows the evolution of the best-fit spectral parameters. It demonstrates that before the 50th day the values of $n_\mathrm{H}$ and the photon index were relatively stable.  

To asses a contribution from harder X-ray radiation, we follow the procedure of \citet{meshcheryakov18}. 
We have downloaded daily light curves of Aql\,X-1 from Swift/BAT Hard X-ray Transient Monitor archive website\footnote{\url{swift.gsfc.nasa.gov/results/transients/index.html}}. For counts-to-flux conversion, we assume that  1 Crab equals to 0.220 counts s$^{-1}$ cm$^{-2}$ in the $15-50$~keV band.
Fig.~\ref{fig:flux_evolution_ergs} clearly indicates that before the $\sim$ 50th day after the peak the X-ray flux is dominated by the soft component. 


\subsection{Optical, UV, and IR data}\label{ss.optical}
The Swift/UVOT observations during the whole period of Aql~X-1 accretion outburst of 2013, in $V$, $B$, $U$, $UVW1$, $UVW2$, and $UVM2$ are shown in  Fig.~\ref{fig:flux_evolution_counts}. 
Errors are purely statistical and correspond to 1-$\sigma$ confidence level.
For the data reduction, images initially preprocessed at the Swift
Data Center at the Goddard Space Flight Center were used.
Subsequent analysis was done following procedure described at the web-page of UK Swift Science Data Centre\footnote{\url{www.swift.ac.uk/analysis/uvot/index.php}}. 
Namely, photometry was performed using {\tt uvotsource} procedure with the source apertures of radius 5 and 10 arcsec for the background for all filters.
The  5 arcsec aperture contains flux from the group of faint stars, located nearby to the Aql X-1 optical counterpart. 


The 1.3-m telescope at Cerro Tololo (Chile) monitored  Aql\,X-1  in $R$ and $J$ bands on a regular basis. We use a publicly available\footnote{\url{www.astro.yale.edu/smarts/xrb/home.php}} SMARTS light curves in our analysis. 
The photometric reduction procedure was performed by the Yale SMARTS XRB team, following the reduction steps described in \citet{buxton12}.

Using  zero points of the passbands, which can be found in \citet{meshcheryakov18}, and the optical extinction to \hbox{Aql\,X-1} $E_{B-V} = 0.64 \pm 0.04$  (see Appendix \ref{sss.redening}), the magnitudes were converted to the flux density units. 
They are compared to  modeled  $\mathcal{F}_\nu$, calculated at 
$\nu = c/\lambda_{\rm eff}$, where $\lambda_{\rm eff}$ are the effective wavelengths~\citep{meshcheryakov18}.

\begin{figure*}
    \centering

\includegraphics[width=0.95\textwidth]{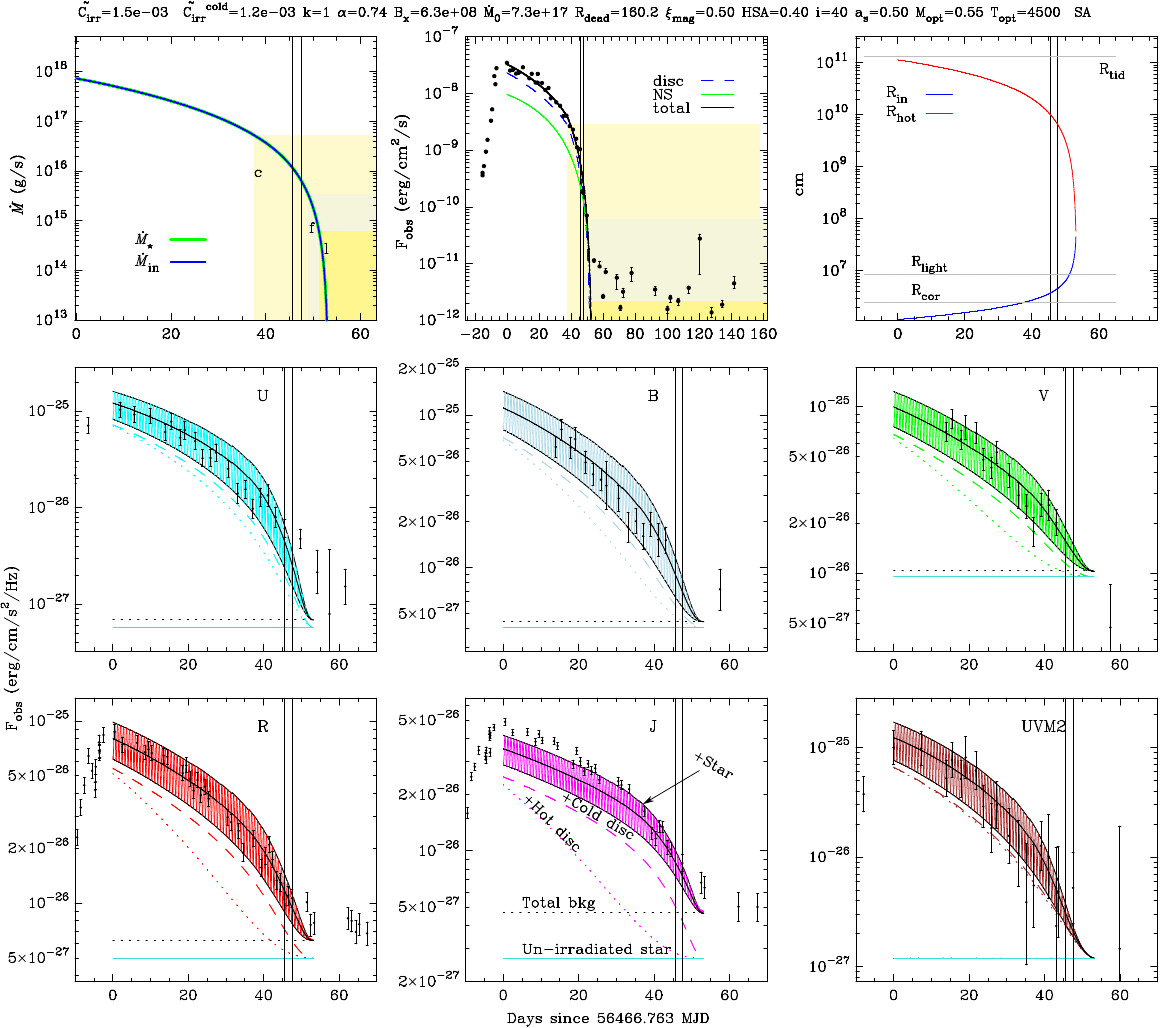}
    \caption{Model \SAmain and observed light curves. The fit parameters are shown at the top and in Table~\ref{tab:result_parametersshort}.  From left to right, from top to bottom, the panels are:  accretion rate, unabsorbed X-ray flux, disc radii, deredened {\vtri flux density at $\lambda_{\rm eff}$} of $U$, $B$, $V$, $R$, $J$, and  $UVM2$. 
    In the 1st and 2nd panel, the colored areas mark stages of evolution (see description in the text, \S\ref{ss.refmodel}).
    In the 3rd panel, the horizontal lines show the corotation radius, light-cylinder radius, and tidal radius. In the panels with the optical flux, a  dotted curve is the hot disc emission, a  dashed curve is the hot$+$cold disc emission. The oscillating curves additionally include the orbit-modulated light from the optical companion. The horizontal dotted line is the background level, which is the sum of the flux from the secondary star during quiescence (calculated in the model; the solid  horizontal line) and the sky background.}
    \label{fig.main_fit}
\end{figure*}
\begin{figure*}
    \centering
\includegraphics[width=0.95\textwidth]{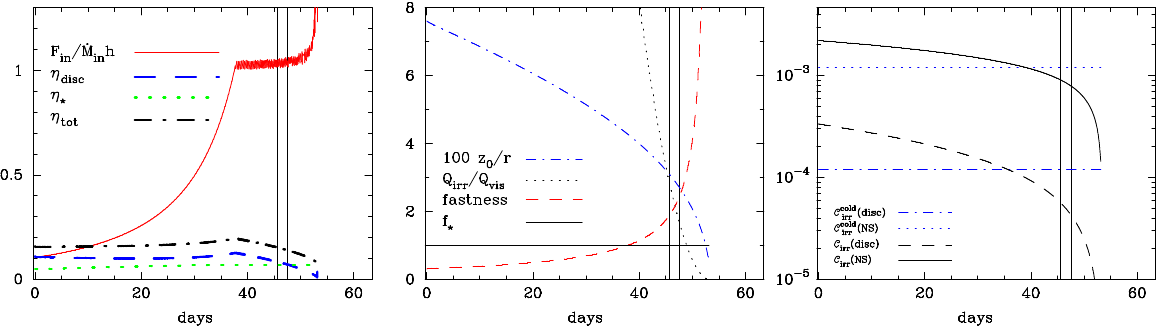}
    \caption{Evolution of key dimensionless values for model \SAmain, see Fig.~\ref{fig.main_fit}. In the left panel: solid line is the 
    modulus of the ratio of the spin-down to spin-up torque {\vsix acting on the neutron star (as prescribed by} Eq.~\eqref{eq.NStorque}); other curves show accretion efficiency in the disc and at the neutron star, and their sum. In the middle panel:  relative semithickness of the hot disc at $\Rhot$, multiplied by 100,  {\vfour irradiation-to-viscous-heat ratio at $\Rhot$,} fastness parameter $\omega_s$,  and ratio $f_\star$ of the accretion rate onto the star's surface to the accretion rate in the disc at $\Rin$. The right panel: irradiation parameters $\mathcal{C}_\irr$ and $\mathcal{C}_\irr^{\rm cold}$ defined by \eqref{eq.Cirrboth}; in the brackets the source of X-ray radiation is indicated.}
    \label{fig.aux_pars}
\end{figure*}
\begin{figure}
     \centering
\includegraphics[width=0.45\textwidth]{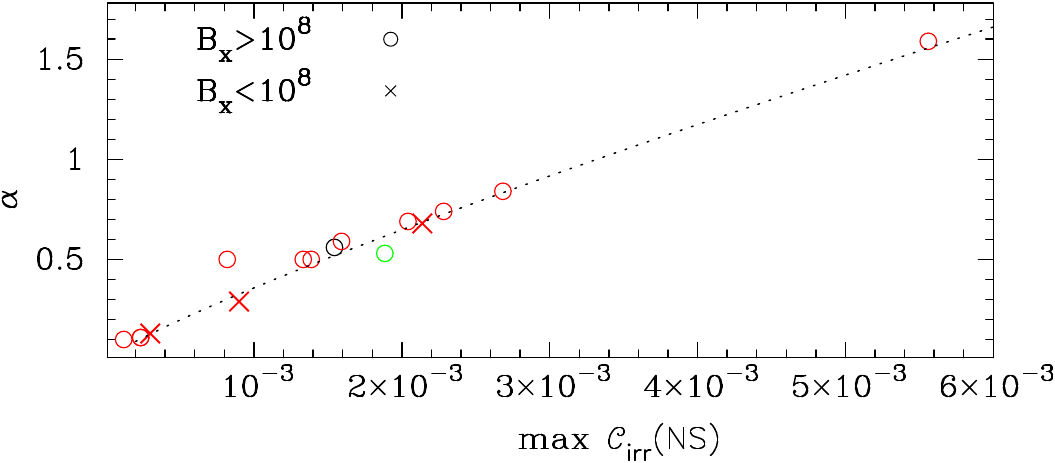}
    \caption{Relation between $\alpha$ and maximum irradiation factor that parametrizes the reprocessed X-ray flux emitted by the neutron star for the models listed in Table~\ref{tab:result_parametersshort}.  Red color marks the \NS scenario, black \PONS, and  green \PO.
    }
    \label{fig.alpha_Cirr}
\end{figure}
\begin{figure*}
     \centering
\includegraphics[width=0.44\textwidth]{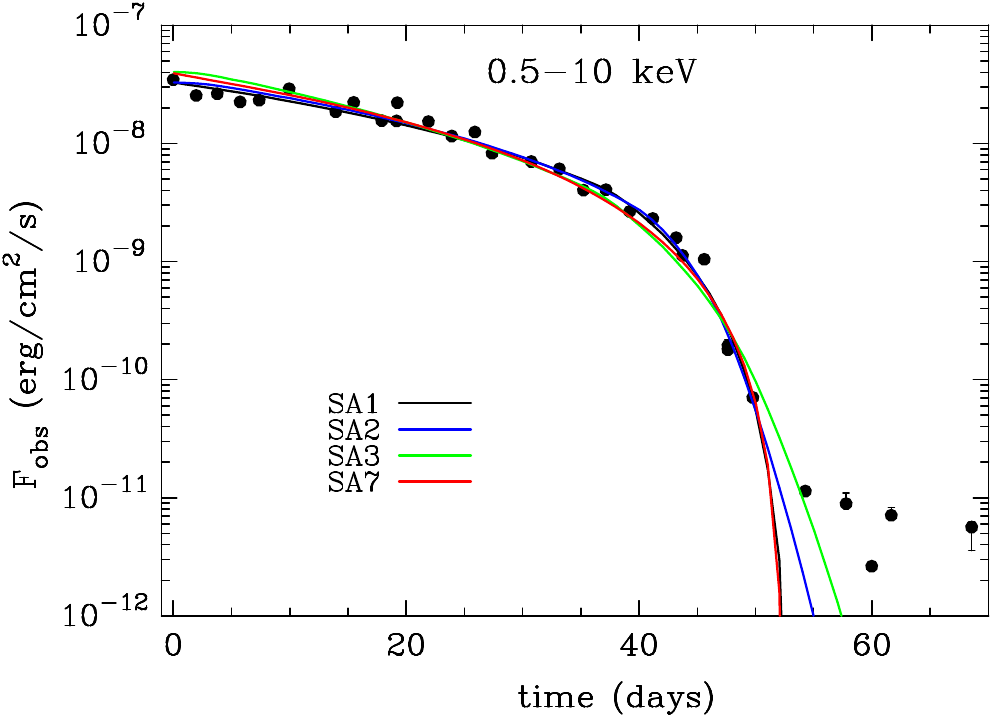}
\includegraphics[width=0.44\textwidth]{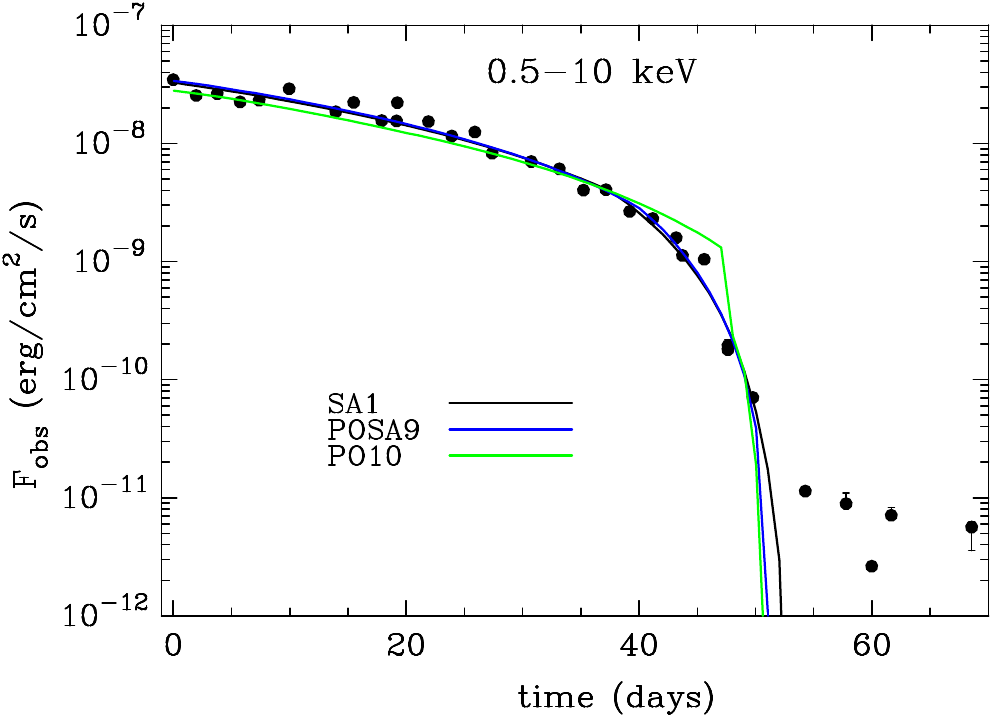}
\includegraphics[width=0.44\textwidth]{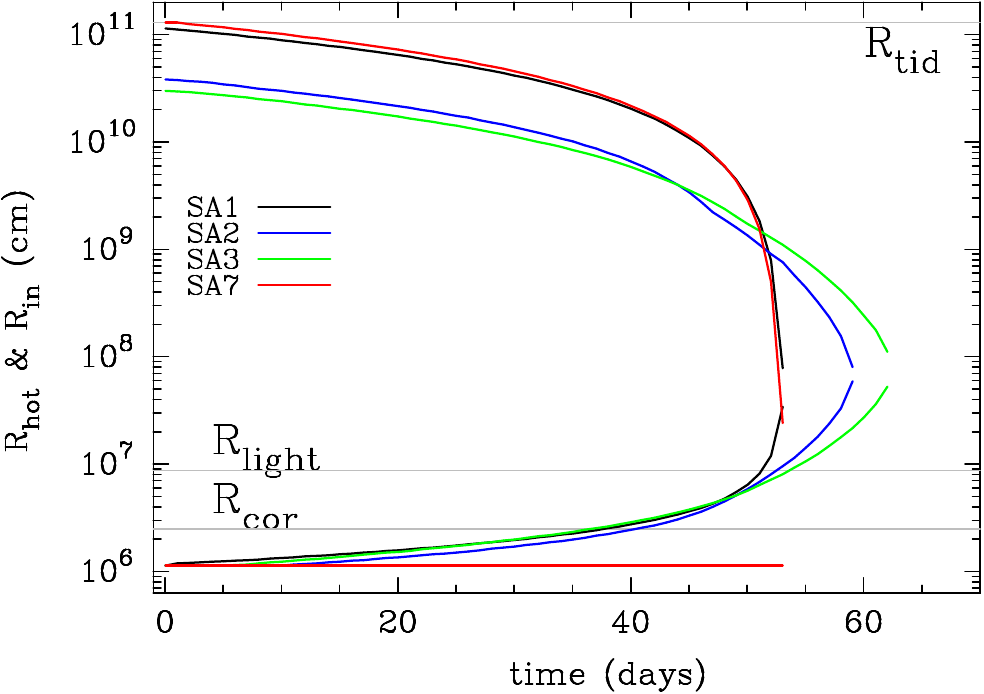}
\includegraphics[width=0.44\textwidth]{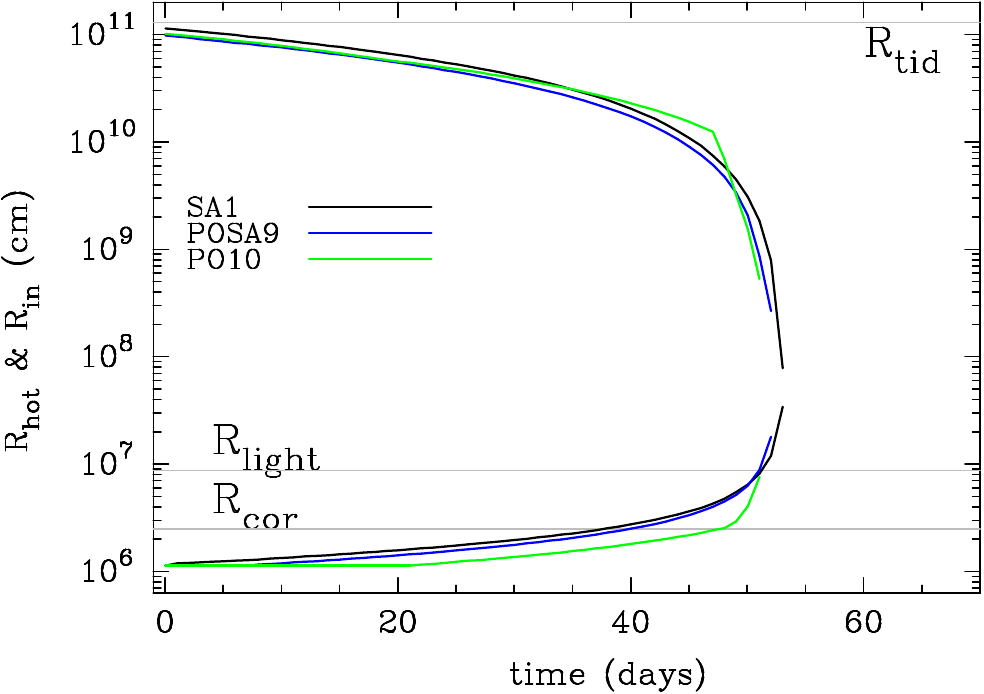}
    \caption{Comparison of different scenarios. {\vsix Left: Effect of self-irradiation. Models \SAmainSmallAlpha and \SAmaink0 have $\Cirr$ about an order less than \SAmain and \smallB do. Right: Effect of different outflow prescriptions listed in Table~\ref{tab:model_scenarios}.}   Top: X-ray light curves. Bottom: the disc inner  and hot-zone radii. The horizontal lines are, from the highest to the lowest: the tidal, the light-cylinder, and the corotation radius.}
    \label{fig.Xcomp}
\end{figure*}
\begin{figure}
    \centering
    \includegraphics[width=0.47\textwidth]{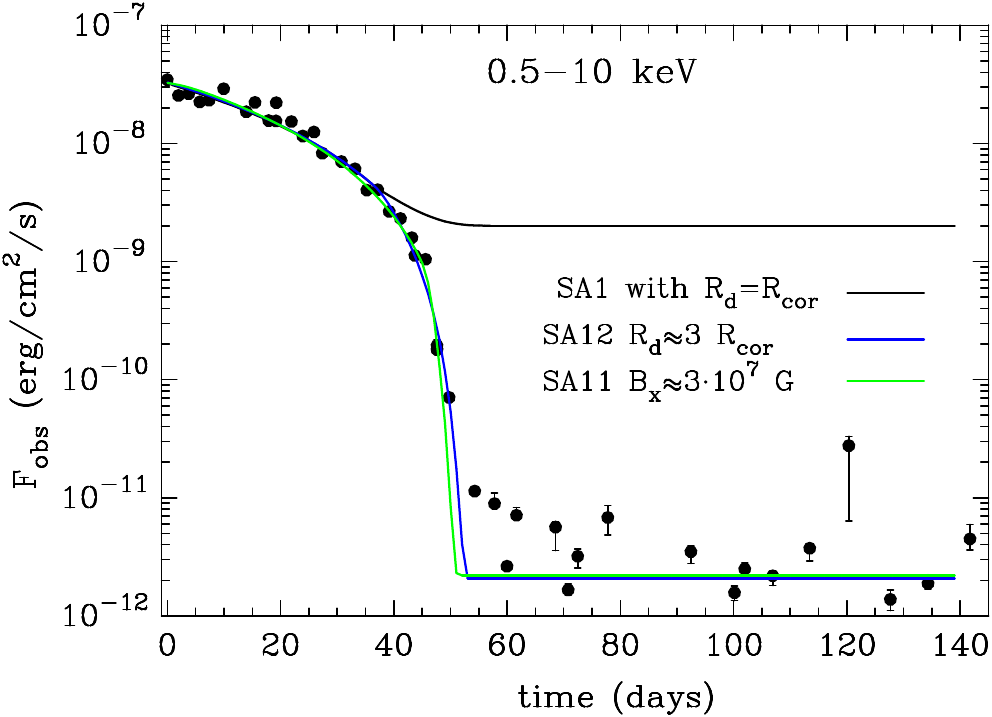}
     \includegraphics[width=0.44\textwidth]{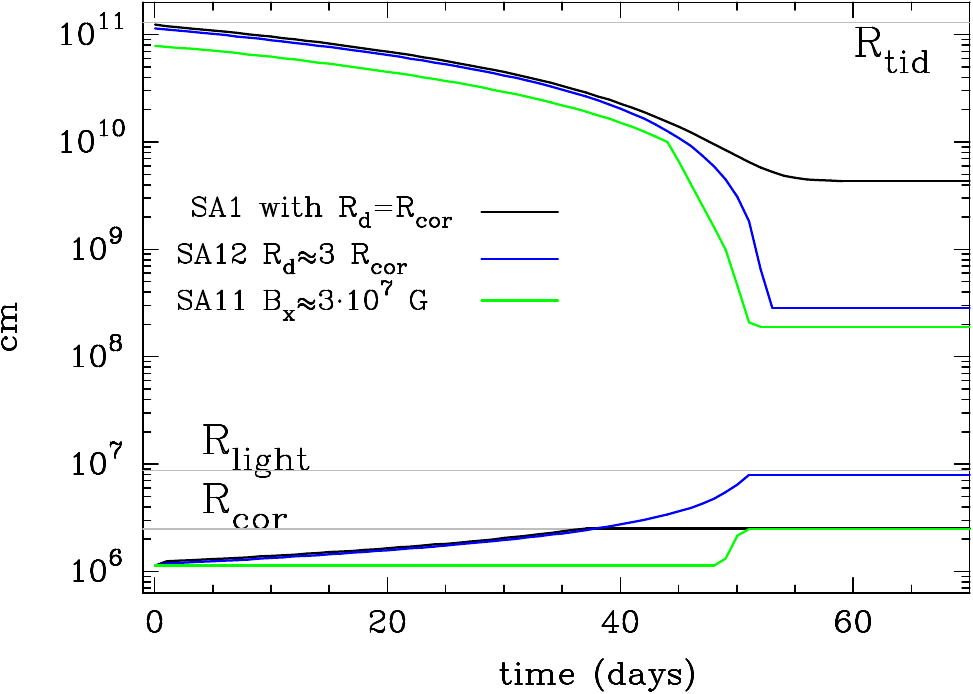}
      \includegraphics[width=0.45\textwidth]{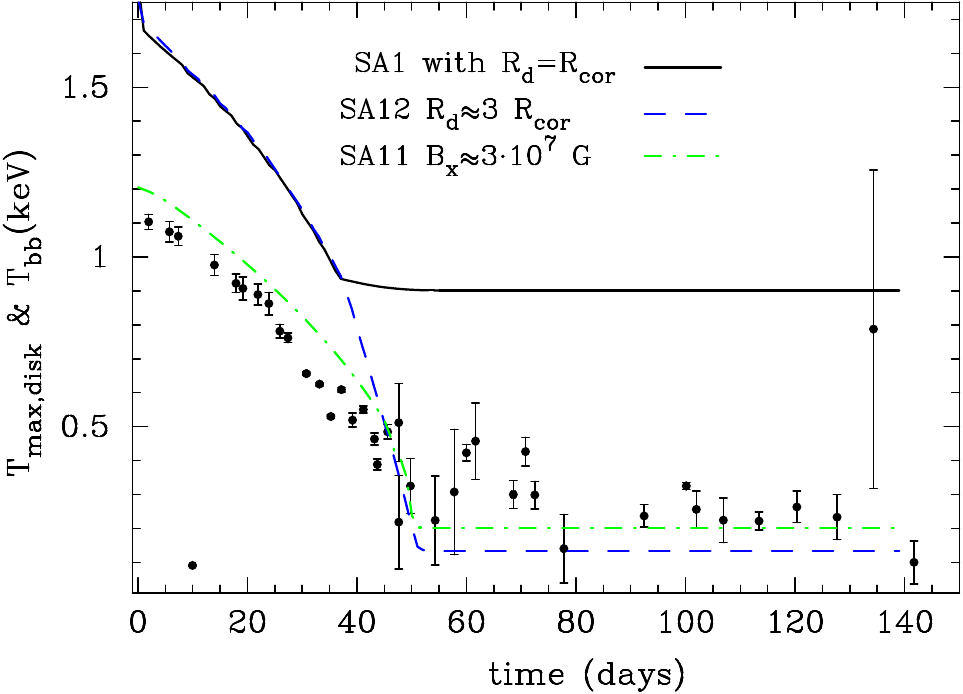}
    \caption{The models with the inner disc radius kept constant when reaching a set value. Two upper panels are similar to Fig.~\ref{fig.Xcomp}. The lower panel shows the maximum effective temperature of the disc (curves) and the  black-body temperature $T_{\rm bb}$ from the spectral fits (dots; see \S\ref{ss.XRT}). }
    \label{fig.plato}
\end{figure}
\begin{figure}
     \centering
\includegraphics[width=0.44\textwidth]{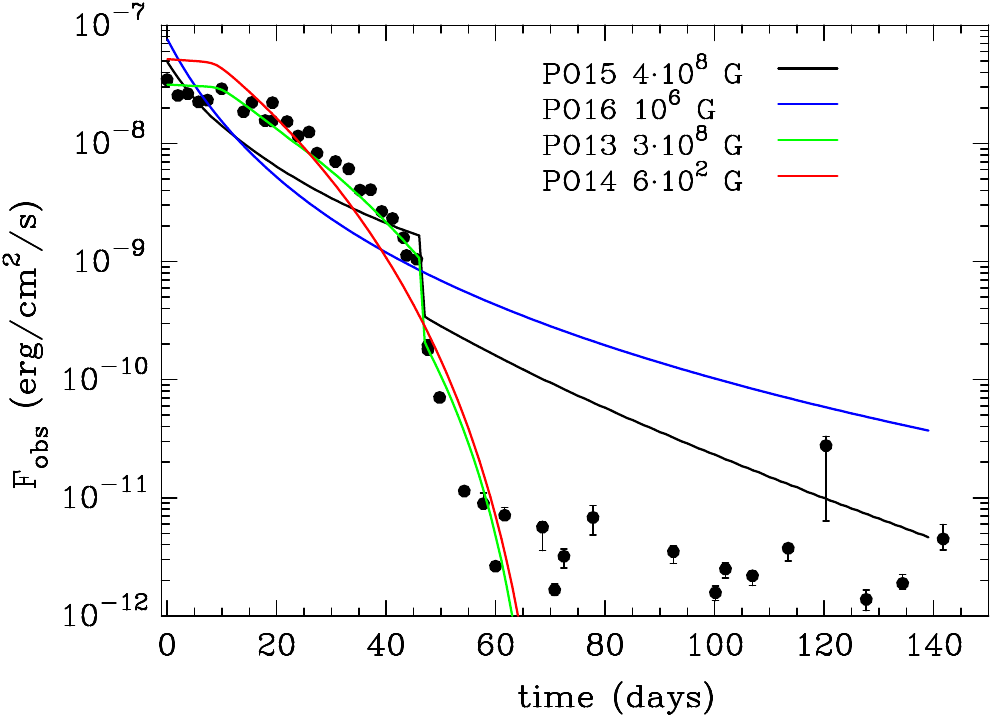}
    \caption{X-ray light curves resulted  in the  models without irradiation of the disc (rejected models).}
    \label{fig.no-irr}
\end{figure}
\begin{table*}
    \centering \caption{Parameters of resulting models.  Columns are: (1)
    Figure number; (2-6) Resulting parameters; (7) Parameter restrictions or
    changes to values given in Table~\ref{tab:AqlX1_pars}; (8-10) Resulting $\chi^2$ and
    $\cod$ for X-ray and optical data; (11) Model ID.  For the `POSA'
    scenario, the material with $v<v_{\rm esc}$ falls onto NS and there is
    no outflow when the disc reaches the star surface.}
    \label{tab:result_parametersshort}
    \begin{tabular}{c|c|c|c|c|c|c|c|c|c|c}
    Fig & $B_x$ (Gs) &
     $\dot{M}_0$ (g/s) & $\alpha$  & $\widetilde{\mathcal{C}}_{\rm irr}$ & $\widetilde{\mathcal{C}}_{\rm irr}^{\rm cold}$  & Comment & $\chi^2$ (X) & $\cod(X)$ & $\chi^2$ (opt) & Model ID \\
     \hline 
    
\multicolumn{10}{c}{{\bf No outflows}}  \\
\ref{fig.main_fit} &   6.3e+08 &   7.3e+17  & 0.74   &   1.5e-03  &   1.2e-03    & $\alpha\geq0.5$ &  21348 & 0.986 & 155 & SA1 \\ 
SF\ref{fig.noutflow_X_U_B_V_R_ae_fac_newt_X14_fcf_alp_small} &   5.2e+08 &   7.9e+17  & 0.11   &   1.3e-04  &   1.6e-03    & $\alpha\leq0.5$ &  21395 & 0.988 & 186 & SA2 \\ 
SF\ref{fig.noutflow_X_U_B_V_R_fitCirr_BX_ae_fac_newt_X14_00} &   6.2e+08 &   9.6e+17  & 0.10   &   1.2e-04  &   1.7e-03    & ~$k=0$ &  52435 & 0.971 & 127 & SA3 \\ 
SF\ref{fig.noutflow_X_U_B_V_R_fitCirr_BX_ae_fac_newt_ktdconst_Mprop_vesc_X14_cold0_fcf} &   5.3e+08 &   8.4e+17  & 0.69   &   1.3e-03  &   0.0e+00    & ~fixed~$\Ccold=0$ &  20805 & 0.986 & 306 & SA4 \\ 
SF\ref{fig.noutflow_X_U_B_V_R_fitCirr_BX_ae_fac_newt_ktdconst_Mprop_vesc_X14_cold0_opt0_fcf} &   7.0e+08 &   9.6e+17  & 1.59   &   3.8e-03  &   0.0e+00    & ~fixed~$\widetilde{\mathcal{C}}_{\rm{irr}}^{\rm{cold}}=0$,~$a_{\rm{opt}}=1$ &  35387 & 0.982 & 571 & SA5 \\ 
SF\ref{fig.noutflow_X_U_B_V_R_fitCirr_BX_ae_fac_newt_ktdconst_Mprop_vesc_noMopt_X14_fcf} &   5.3e+08 &   7.9e+17  & 0.11   &   1.3e-04  &   3.8e-03    & ~$a_{\rm{opt}}=1$ &  21406 & 0.988 & 193 & SA6 \\ 
SF\ref{fig.noutflow_X_U_B_V_R_fitCirr_Bx_ae_fac_newt_X14_alp_big_fcf} &   1.0e+02 &   1.1e+18  & 0.68   &   1.3e-03  &   5.5e-04    & fixed~$B_x$,~$\alpha\geq0.5$ &  34573 & 0.979 & 162 & SA7 \\ 
SF\ref{fig.noutflow_X_U_B_V_R_fitCirr_smallBx_ae_fac_newt_X14_alp_small_fcf} &   1.0e+02 &   9.8e+17  & 0.13   &   1.6e-04  &   1.2e-03    & fixed~$B_x$,~$\alpha\leq0.5$ &  32725 & 0.981 & 177 & SA8 \\ 
\multicolumn{10}{c}{{\bf Partial outflow if~$\Rin>\Rcor$}}  \\
SF\ref{fig.romanova2018_X_U_B_V_R_ae_fac_newt_X14_fcf} &   5.5e+08 &   8.1e+17  & 0.56   &   9.7e-04  &   1.4e-03    & ~ &  20469 & 0.986 & 175 & POSA9 \\ 
\multicolumn{10}{c}{{\bf Complete outflow if~$\Rin>\Rcor$}}  \\
\ref{fig.Xcomp} &   3.6e+08 &   7.1e+17  & 0.53   &   1.2e-03  &   1.0e-03    & ~ &  49943 & 0.967 & 249 & PO10 \\ 
\multicolumn{10}{c}{{\bf Dead disc for a plato, which is one point in X:}}  \\
\ref{fig.plato} &   3.3e+07 &   9.0e+17  & 0.29   &   5.2e-04  &   1.1e-03    & $B_x$~explains~plato,~$R_{\rm{dead}}=\Rcor$ &  21647 & 0.948 & 163 & SA11 \\ 
\ref{fig.plato} &   6.3e+08 &   7.3e+17  & 0.74   &   1.5e-03  &   1.2e-03    & $R_{\rm{dead}}\approx3.3\Rcor$~explains~plato~in~\SAmain &  21348 & 0.986 & 155 & SA12 \\ 
\multicolumn{10}{c}{{\bf No disc irradiation:}}  \\
\ref{fig.no-irr} &   3.3e+08 &   8.2e+17  & 0.06   &   1.0e-10  &   0.0e+00    & fixed~$\widetilde{\mathcal{C}}_{\rm{irr}}=\widetilde{\mathcal{C}}_{\rm{irr}}^{\rm{cold}}=0$~and~$B_x$; &  55158 & 0.981 & 1574 & PO13 \\ 
\ref{fig.no-irr} &   6.0e+02 &   1.5e+18  & 0.06   &   1.0e-10  &   0.0e+00    & fixed~$\widetilde{\mathcal{C}}_{\rm{irr}}=\widetilde{\mathcal{C}}_{\rm{irr}}^{\rm{cold}}=0$; &  334770 & 0.874 & 1250 & PO14 \\ 
\multicolumn{10}{c}{{\bf No irradiation and constant outer radius $\Rhot=R_{\rm{tid}}$}}  \\
\ref{fig.no-irr} &   4.2e+08 &   1.3e+18  & 4.00   &   1.0e-10  &   0.0e+00    & fixed~$\widetilde{\mathcal{C}}_{\rm{irr}}$,~$\widetilde{\mathcal{C}}_{\rm{irr}}^{\rm{cold}}$~\&~$R_{\rm{hot}}$ &  423694 & 0.830 & 2137 & PO15 \\ 
\ref{fig.no-irr} &   1.0e+06 &   2.2e+18  & 4.00   &   1.0e-10  &   0.0e+00    & fixed~$\widetilde{\mathcal{C}}_{\rm{irr}}$,~$\widetilde{\mathcal{C}}_{\rm{irr}}^{\rm{cold}}$,~$B_x$~\&~$R_{\rm{hot}}$ &  907624 & 0.622 & 1990 & PO16 \\ 
\multicolumn{10}{c}{{\bf Irradiation of hot disc with constant outer radius $\Rhot=10^{10}~$cm, no cold disc}}  \\
SF\ref{fig.corblock_X_U_Cirrcold0_Rhot1e10_X14_Bx338} &   3.3e+08 &   3.6e+18  & 0.07   &   1.0e-10  &   2.4e-03    & fixed~$\widetilde{\mathcal{C}}_{\rm{irr}}^{\rm{cold}}$=0~and~$B_x$ &  969733 & 0.717 & 3212 & PO17 \\ 
\multicolumn{10}{c}{{\bf Irradiation of whole disc; hot part has constant outer radius}}  \\
SF\ref{fig.corblock_X_U_Rhot1e10_X14_Bx338} &   3.3e+08 &   3.4e+18  & 0.07   &   5.0e-02  &   1.3e-05    & fixed~$B_x$,~$\Rhot=10^{10}$~cm &  966438 & 0.717 & 2121 & PO18 \\ 
SF\ref{fig.corblock_X_U_Rhottid_X14_Bx338} &   3.3e+08 &   3.3e+17  & 4.00   &   1.7e-03  &   1.2e-05    & fixed~$B_x$,~$\Rhot=R_{\rm{tid}}$ &  1528271 & 0.512 & 1823 & PO19 \\ 
\hline \end{tabular}
\end{table*}

\section{Results}\label{s.results}

{\vfour We have performed numerous fits to the observed light curves of the 2013 burst decay of Aql X-1
by the model light curves produced with the modified \freddi-code. For fitting, the data in $0.5-10$~keV and $U$, $B$, $V$, and $R$ in the interval between MJD\,56465.763 and MJD\,56516.533 (between the peak and 53th day after the peak) is used.}
Resulting models are plotted for a longer time interval, see, for example, Fig.~\ref{fig.main_fit} and figures in the Supplement Material, {\vfour and also for $J$ and $UVM2$ data.}
Table~\ref{tab:result_parametersshort} summarizes essential results.
{\vfour For models \SARcor and \SARdead, we fit data before the 135th day. In this case all X-ray points after the 53th day are substituted by one point, obtained from the spectrum integrated in the interval MJD\,56550-56650 -- the point with a long horizontal bar in Fig.~\ref{fig:sppar_evolution}. In the rest of the models, down starting from \irrI,  we fit the data in interval  MJD\,56465.763-56530.763 (before the 65th day).}

Our fitting procedure provides estimates  of the following parameters: {\vfour the peak accretion rate,}  $\alpha$-parameter in the hot disc,  irradiation parameter $\mathcal{C}_{\rm irr}$,  magnetic field of the neutron star, whereas we fix value of $\xinner$ and $a_{\rm opt}$.
These parameters are interrelated. 
In particular:
\begin{itemize}
    \item[]  $\alpha$ and $\mathcal{C}_{\rm irr}$ affect the overall rate of the source evolution;
\item[]   $B_x$ and $\xinner$ affect the knee  position;
 \item[]   $\mathcal{C}_{\rm irr}$, $\mathcal{C}_{\rm irr}^{\rm cold}$, and $a_{\rm opt}$ affect
the optical to X-ray flux ratio.
 \end{itemize}

A performance of a particular model  can be assessed by its $ \chi^2$ or $\cod$ (the coefficient of determination\footnote{$\cod = 1- \sum(y_i - m(t))^2/\sum(y_i - \overline{y})^2$ measures how well observed data are replicated by the model, where $y_i$ is the data, $m(t)$ is the model, $\overline{y}$ is the mean of $y_i$. The model prognosis is perfect when $\cod= 1$.}). For the optical data ($U$, $B$, $V$, and $R$), values of $\chi^2$ are calculated for  the orbit-averaged  total flux.
{\vfour The statistics are calculated for interval 0--53~d for all fits; there are 27 X-ray and 81 optical points in in the interval.}

The nature of the data is so that a lot of local minima of the fit statistic exist. 
Thus it is impossible to circle out a single best model.
Still, let us consider a particular model in order to compare others to it.

\subsection{A particular model}\label{ss.refmodel}

In Figure~\ref{fig.main_fit} model \SAmain is presented when  the shrinking of the hot zone drives the X-ray flux drop around the 
35th day.
The inner radius of the disc levels with the corotation radius at 
37.45 day
at $\dMin \approx 5.3\times 10^{16}$~g/s when $\Rhot\approx 2\times 10^{10}$~cm and the bolometric luminosity of the star and disc is  $9.2 \times 10^{36}$~erg/s for the total accretion efficiency $\eta_{\rm}\approx 0.19$. 
In the 1st and 2nd panel of Fig.~\ref{fig.main_fit}, the shaded rectangles show the stages when the inner disc radius is larger than the corotation radius $\Rcor$ (rectangle marked with letter 'c'); when the inner disc radius is larger than the light-cylinder radius $R_{\rm light}$ (letter 'l'); when the irradiation-to-viscous flux ratio $Q_{\rm irr}/Q_{\rm irr} < 1$ at $R_{\rm hot}$ and the radius of the hot disc $R_{\rm hot}$ is found using the front velocity, as described in \S2.4 (letter 'f').
    
In this particular model, no outflow is present, and the accretion onto the neutron star should continue after 
37.45 day in a sporadic fashion. 
The low-level X-ray plato after the 
55th day is not addressed in this model \footnote{
Note that whenever the inner disc radius becomes greater than the radius of the light cylinder, the material of the disc is dispersed by the pulsar wind, {\vseven at least, in its hot part,} and the model solution is purely formal for those times.}.

It is interesting that, contrary to a primary expectation, the X-ray flux detected by an observer is dominated by the disc, which is roughly twice as brighter comparing with the flux from the neutron star. 
This is explained by (i) a rather fast rotation of the star yielding the bolometric luminosity of the disc about 60\% brighter {\vsix than that of the star}; (ii) {\vsix the angular distribution of the disc emission $\Psi(\theta)=2\, \cos \theta$ causes the mild enhancement  at $\theta = i = 40^o$, comparing to an isotropic case}.

Figure~\ref{fig.aux_pars} shows corresponding evolution of some key dimensionless parameters: accretion efficiency (in the disc and on the star); the disc semithickness $z_0/r$,  irradiation-viscous-flux ratio $Q_\irr/Q_\vis$ at the outer radius of the hot zone, the fastness parameter $\omega_s=\omega_\star/\omega_{\rm K}(\Rin)$, the accretion penetration parameter $f_\star=\dot M_\star/\dMin$; the irradiation parameters applicable to the hot/cold disc  for the flux from disc/star. 
Note that irradiation parameter for the cold disc does not vary since its semithickness is fixed.

\subsection{Turbulence parameter and irradiation parameter}

To begin with, we set a high/low limit for a free parameter $\alpha$ at 0.5 (keeping other settings the same).
It turns out that the resulting models behave similarly and yield comparable fit statistic (see  models SA1 and SA2, Fig.~\ref{fig.Xcomp}, left panels). 
It can be seen (Table~\ref{tab:result_parametersshort}) that the rate of evolution in discs with different $\alpha$ is the same  because the irradiation parameter is properly adjusted.

Fitting X-ray data alone, a relation between $\alpha$ and $\Cirr$ can be found. 
This relation between the maximum irradiation parameter $\Cirr$ (i.e., at the peak of the burst, see \eqref{eq.Cirrboth}) and $\alpha$ for models from Table~\ref{tab:result_parametersshort} is shown in  Fig.~\ref{fig.alpha_Cirr}.  
The degeneracy between them can be partly resolved by determining $\Cirr$ from the optical data. 
Ignoring optical contribution from other sites (the cold disc and  optical companion) results in a larger hot disc.
Herewith, the X-ray light curve fit becomes worse by almost two times and estimated $\alpha>1$ (model \SAmainNoColdStar). 

{\vfour It should be noted here that the $\alpha$--parameter, found by fits, is model-dependent. 
First, a method to estimate $\alpha$ from the dynamical evolution of the disc, ignoring winds from the disc surface that speed up the evolution, overestimates $\alpha$. 
Second, $\alpha$ depends on the hot-zone radius definition, which is rather simple in the present version of our scheme, see \S\ref{s.Rout}.}

For the models with $\alpha>0.5$, the value of $\alpha_{\rm cold}$ (we take 0.01) does not affect the fits, since the irradiation controls practically the whole time interval used for fitting. 
This value affects the velocity of the cooling front and affects fitting results if $\Cirr$ and $\alpha$ are relatively small.

We also test the properties of irradiation. 
Model \SAmaink0 (Fig.~\ref{fig.Xcomp}, the left panels) has the irradiation parameter that does not depend on the disc thickness ($k=0$ in \eqref{eq.Cirrboth}) and its X-ray fit  is about 2.5 times worse than that for \SAmain. 
Thus the variable irradiation parameter is favoured by our analysis (see the right panel of Fig.~\ref{fig.aux_pars}). 
This variability is a geometrical effect, since  it ensues from  changing disc  thickness.

\subsection{Optical flux}

{\vsix In our model, not only the hot ionized disc, but also the outer cold ring and the optical companion can contribute to the optical flux of the source.}
The total optical flux is calculated as the sum of the  source  flux  and a `sky background'. 
The sky background  is calculated  as the quiescent optical flux~(\citealt{meshcheryakov18}; table 4) minus the flux of the non-irradiated optical star. 
The flux of the non-irradiated companion depends on the chosen  parameters of the binary ({\vsix temperature $T_{\rm opt}$, binary semi-axis $a$, mass ratio $q$, filling factor $R_{\rm pol} / R_{\rm pol}^{\rm Roche} $, and inclination $i$, see  Table~\ref{tab:AqlX1_pars}). 
For chosen parameters, radius of the optical star' is about $0.3\,a$.}
{\vsix Due to the orbital variation of the visible area of the irradiated surface of the star,}
the resulting optical light curves look very dense {\vsix(as in two lower rows in Fig.~\ref{fig.main_fit})} .

At the end, such modeling of optical data turns out to be partly successful. 
On one hand, the level and  trend of the optical flux are explained better with the inclusion of the cold disc and the optical star than without. 
This inference can be made on inspecting the optical $\chi^2$-values in Table~\ref{tab:result_parametersshort} and the optical light curves in the Supplement Material (compare models \SAmainNoCold, \SAmainNoColdStar,  and \SAmainNoOpt with model \SAmain). 
On the other hand, we could not explain fully the oscillations of the optical data by the orbit-modulated flux from the optical star. 
This is maybe partly due to the fact that, in addition to period-modulated fluctuations,  there are fluctuations in the illuminating  X-ray flux (similar to those seen in the X-ray curve), not accounted for in our modeling. 
Also, the amplitude of the model optical fluctuations is generally larger than observed. 
This may indicate that the albedo value 0.5 is an underestimation, or that the shadow on the star is actually bigger than what the model disc produces.

\subsection{Magnetosphere-disc interaction}

{\vsix Concerning the resulting light curves, two scenarios, the one without outflows (\NS) and the one that adopts results of MHD simulations by \citetalias{Romanova+2018} (\PONS),  provide very similar results (compare models \SAmain and \POSAmain). The latter describes a gradual propeller turn-on, and
the similarity is due to the fact that the adopted dependence for a 
propelled mass portion  $f_{\rm eff}(\omega_s)$ holds for outflows with velocity greater than the escape velocity, ensuring the loss of mass from the disc. The dependence on the fastness
has a high power index (about 4, see \S\ref{s.Mdot_ns} and \citetalias{Romanova+2018}) and $f_{\rm eff}$ differs very little from 1  
just after the inner disc edge recedes beyond the corotation radius, before the 50th day, see Fig.~SF\ref{fig.romanova2018_X_U_B_V_R_ae_fac_newt_X14_fcf}. }
Overall, until the moment when the disc edge levels with the light cylinder radius, only about $10^{20} $~g  is propelled away in model \POSAmain. 

Comparing different scenarios of accretion inhibition near the corotation radius (Fig.~\ref{fig.Xcomp}, the right panels), we see that a model with a very efficient propeller (\POmain) fits worse X-ray data because it produces a too dramatic drop, which is not observed.
Thus, a scenario with a gradual decrease of the averaged accretion rate on the neutron star surface, like \NS or \PONS, is preferred. 

Further, we consider models with small magnetic fields by setting an upper limit on $B_x$. 
We find that  their X-ray  fits are generally, and independently from values of $\alpha$ and $\Chot$, worse (models \smallB and \smallBsmalla). 
This is explained by a particular behaviour of the accretion efficiency in the case of a strong magnetic field, which makes a knee {\vsix on an X-ray light curve} more pronounced (compare \SAmain and \smallB in Fig.~\ref{fig.Xcomp}, the left panels).
The rotating {\vsix magnetosphere of the} neutron star adds up to the torque at the inner disc radius and enables the disc-accretion efficiency to have a maximum when $\Rin=\Rcor$ (see \eqref{eq.Fin_1part}, \eqref{eq.eta_disc}, and Fig.~\ref{fig.aux_pars}, the left panel, the blue dashed line).

\subsection{Is there a disc at the end?}\label{s.bestfits}

There  is a possibility that a `left-over' disc remains after an outburst~\citep[indicated, for example, by an observed double burst from ms pulsar IGR J00291+5934, see][]{Hartman+2011}. 
The two flares of the pulsar separated by mere 30 days forced the authors to conclude that  much of the disc was conserved after the 1st burst, apparently because the mass loss virtually stopped.  
This case can be calculated in the \PONS or \NS scenario with a pre-set  value of the final disc radius $\Rdead$. 
A final inner torque value $F_{\rm dead}$ is related to this radius via \eqref{eq.Fdead}. 
When, in the course of viscous evolution, the torque values at the inner radii level with $F_{\rm dead}$, the accretion flow through the inner edge stops. 
Then the emitted energy is provided by a constant inner torque $\Fin$ even when the accretion rate $\dMin$ is zero (see Eq.~\eqref{eq.L_bolometric}). 
A dead disc around a magnetized star produces a constant level of radiation supplied by the energy of the neutron star rotation; the flux depends on $\Rdead$ and $B_x$.
For example, if we set $\Rdead = \Rcor$ in a model with $B_x>10^8$~G, we get a very luminous plato contradicting observations (see Fig.~\ref{fig.plato}, the top curve).
Let us consider other possibilities.

\paragraph*{(a) A dead disc hovering at the corotation radius.} 
$\Rdead = \Rcor$ and $B_x \sim (3-4)\times 10^7$~G, model \SARcor.

A scenario of a dead disc or disc-reservoir extending down to the corotation radius has a long history. It  partly bases on the reasoning that the propeller mechanism cannot launch matter near the corotation radius because this is not energetically self-consistent. 
In a numerical time-dependent model for a disc in a binary system, \citet{armitage-clarke96} set the inner radius at $\Rcor$ and observe it to remain there for a long time, even when $\dot M =0$,  since the mass cannot leave the disc. \citet{Rappaport+2004}  suggest a disc inner edge remains at the corotation radius with a decaying mass flux in order to explain high spin-down rates  and apparent accretion over wide range of $\dot M$ in millisecond AXPs.
\citet{dangelo-spruit2012} suggest that a `trapped' state of the disc is possible when  the disc with an arbitrary low accretion rate `hovers' around the corotation radius.  

In order for the thin disc to emit not too much X-rays when its inner edge is at the corotation radius (about 15 km above the neutron star surface), the magnetic field should be lowered.
In Fig.~\ref{fig.plato} we  present a corresponding fit \SARcor. 
The magnetic field, consistent with such disc luminosity, i.e. to fit the plato, is a decade less than in our reference model, just $\sim(3-4)\times 10^7$~G. 

No matter is propelled away during an outburst in such scenario and the disc's inner radius first approaches, then oscillates around, and, finally, freezes at the corotation radius.
(Though presumed,  oscillations of the inner radius are not resolved in our numerical model.)
It is inevitable that matter should somehow leave the disc through its inner radius until a constant $F$-profile settles on over the entire hot disc because the outer hot disc radius $R_{\rm hot} (t)$ is not increasing (on the contrary, $R_{\rm hot} (t)$ can only move inward {\vsix after the peak of an outburst}).


In this model, the drop of luminosity at 
$\sim 45$th  day happens due to  the fast shrinking of the hot zone. We note  the modeled light curve fits the X-ray data   worse than  the models with higher magnetic field do. 
However, given the schematic nature of all scenarios, this one cannot be excluded.

\paragraph*{(b) A dead disc hovering farther than the corotation radius.} $\Rdead > \Rcor$ and $B_x<7.5\,\times 10^8$~G, model \SARdead.

To match the luminosity level at the plato stage, {\vsix see \eqref{eq.Fin_2part} and \eqref{eq.L_bolometric},}
$$
 L_{\rm plato} = \kappatd \, \frac{\mu^2\, \sqrt{G\, \mns}}{R_{\rm dead}^{9/2}}\, ,
$$
the inner radius of the thin dead disc should be equal to $\Rdead = \Rcor \times (2\, L_{\rm knee}/3\, L_{\rm plato})^{2/9}$, where luminosity is bolometric. 
The magnetic field as in the model \SAmain requires $\Rdead \sim 3\,\Rcor$ for this to happen (model \SARdead, see Fig.~\ref{fig.plato}).  {\vsix Notice that the inner radius is self-consistently less than the light cylinder radius (see Table~\ref{tab:AqlX1_pars}): in the opposite case an accretion flow would be swept away by the pulsar wind. 
That is, in order for this scenario to be feasible at all, the required magnetic field cannot be too high: $B_x \lesssim 6.3\times 10^7 \,\kappatd^{-1/2}\, \, L_{34}^{1/2} \, P_{-3}^{9/4} \, M_{1.4}^{-1/4}\,  R_6^{-3}~ {\rm G}\, 
$.
 For our parameters (Table~\ref{tab:AqlX1_pars}) and esimated luminosity (\S\ref{ss.plato}), if $B_x$ was  $\gtrsim 7.5\times 10^8$~G, the radius $R_{\rm dead}$ should be higher than $R_{\rm light}$ to comply with the plato level.}

Finally, the maximum temperature in the disc, both in  \SARcor and \SARdead, is consistent with the black-body temperature obtained from the spectral fits (Fig.~\ref{fig.plato}). 
One could not expect much accuracy here, since the disc spectrum is different from that of a black body.

\subsection{The shrinking hot disc is reprocessing X-rays}\label{s.badfits}

We obtain that the brightness in the optical bands during the  2013 outburst of Aql\,X-1 can be explained only if the disc zones, emitting in the optical, are heated by the central X-rays.
This is in agreement with the conclusion by \citet{meshcheryakov18} who investigated the rise of X-ray and optical light curves of the same outburst of Aql\,X-1 and obtained following accretion disc parameters around the outburst maximum:
the disc accretion rate $\dot{M}\approx0.66\times{}\dot{M}_{\rm Edd}= 1.29\times10^{18}$~g/s;   the outer disc radius $R_{\rm d}\approx{}R_{\rm tid} = 1.9\times{}R_\odot = 1.3\times10^{11}$~cm; and    the irradiation parameter (responsible for all optical flux) with respect to $0.5-100$\,keV flux     $\mathcal{C}_{\rm irr} \approx10^{-3}$.
The peak accretion rate $\dMin$ obtained in our models with a small magnetic field (for example, \smallB) agrees well with the above value from~\citet{meshcheryakov18} (see Table~\ref{tab:result_parametersshort}). 
{\vsix Our irradiation parameter $\mathcal{C}_{\rm irr}$ agrees with the previous estimate too: it lies in the interval $(2-3.2)\times 10^{-3}$, see Fig.~\ref{fig.alpha_Cirr}.}

Importance of irradiation and hot-disc size change can be illustrated by a set of unsuccessful models: \irrI and below in Table~\ref{tab:result_parametersshort}.
Supplement Figures~SF\ref{fig.corblock_X_U_Cirr0_Bx3e8_ae_X14} and SF\ref{fig.corblock_X_U_Cirr0_fitBx_ae2_X14} show the light curves obtained  when the disc is non-irradiated (models \irrI and \irrII, respectively).
Accretion rate evolution in models with a constant-size hot zone (models \irrIII--\irrX in Table~\ref{tab:result_parametersshort}),
fails to explain the X-ray light curve (see Figure~\ref{fig.no-irr} and supplement
Figs.~SF\ref{fig.corblock_X_U_Cirr0_fitBx_Rconst_ae_X14}-- SF\ref{fig.corblock_X_U_Rhottid_X14_Bx338})\footnote{We choose \PO{} scenario here because, among all scenarios, it produces steeper drops of the flux near the knee.}.  
Thus, the `exponential' stage of the disc evolution  in the 2013 burst of Aql\, X-1 did not occur or was very short.

\section{Discussion}\label{s.disc}
\subsection{Comparison to other models}
  
The disc viscous evolution around magnetized stars drew attention of many researches.
Evolution of a disc at the brink of the two regimes, accretion and propeller, was studied by \citet{Spruit-Taam1993}. 
They set a constant inflow accretion rate and considered oscillations of the disc structure when its inner radius was shifting around $\Rcor$. 
They have set a special boundary value on the viscous torque, increasing it with increasing  $\omega_s$ and representing  a ``centrifugal barrier''  of an exponential form for $\omega_s>1$.
This disc with such inner conditions is realized in our code by setting the final inner radius of the disc $\Rdead=\Rcor$. 
Oscillations of the inner edge are not resolved in our code, {\vsix because} it is focused on the global viscous evolution.
    
\citet{dangelo-spruit2010} show  that if the accretion rate is slightly less than {\vsix some} critical accretion rate, the time scale of the oscillations is of order of the viscous time scale calculated at $\Rcor$, and the oscillations look like sinusoidal modulations. The less the incoming accretion rate, the less frequently  occur the `overflows'.  \citet{dangelo-spruit2010} suggest that 1-Hz QPO demonstrated in the decline phase by the AXMP SAX\,J1808.8-3658 
 in  several outbursts~\citep{Patruno+2009}, and by the AXMP NGC 6440 X-2,
 are manifestations of such oscillations near the corotation radius.  There is  also a  0.5~Hz QPO seen in AMXP IGR J00291+5934~\citep{Hartman+2011}.
\citet{dangelo-spruit2010} conclude that oscillations can continue for any low value of the accretion rate.

 It is possible that during the decay of an outburst the inner disc radius, when reaching the corotation radius, continues to recede farther in an oscillatory manner. 
Such character of the flow was also demonstrated in numerical simulations (see references in \S\ref{s.model}).
However, in an evolving disc the mass inflow decays since the mass of the hot disc diminishes.  
As a terminal {\vsix state},  the torque value at the inner edge of the disc becomes equal to  the torque value at the outer edge of the hot disc:  a truly `dead' disc configuration emerges~\citep{lipunova2015}. 
There is no movement of matter along the radius. The accretion rate is zero but the disc radiates the energy supplied by the neutron star's spin-down, in accordance with Eq.~(\ref{eq.L_bolometric}) (see also \S\ref{ss.plato}).

{\vsix \citet{armitage-clarke96} witnessed a formation of a disc-reservoir, limited by $\Rcor$ and $R_{\rm tid}$, which decelerated the central star by transferring the angular momentum to the orbital motion.  In their  time-dependent  code for a disc around a magnetized \hbox{T Tauri} star, they forbid accretion for $\Rin>\Rcor$.
The outer boundary condition for their disc in a binary was the same as ours: a zero radial velocity; the mass of the disc was set as an initial condition and the inner disc edge was set just beyond  $\Rcor$. 
Differently from us they set the inner boundary condition $\Sigma=0$ at $\Rcor$ and considered the magnetic torque that was distributed over radius.
They have demonstrated an enhanced disc luminosity in a presence of the magnetic torque, which our model visibly reproduces. 
}

{\vsix Previously,} it has been suggested that the knee position can be used to estimate the magnetic field strength of the star. 
It can be done if one assumes a very efficient blockage of accretion onto a star, like in our scenario \PO.
Equating $\Rcor = \xinner \, R_\mag$, using definitions \eqref{eq.rmag} and \eqref{eq.Rcor}, one gets
$$
 B_x \approx 4.3\times 10^7\xinner^{-7/4}\, \eta_{0.2}^{-1/2}\, L_{36}^{1/2} \, P_{-3}^{7/6} \, M_{1.4}^{5/6}\,  R_6^{-3}~~ {\rm G}\, 
 $$
 or $B_x \approx 3.6 \times 10^8$~G for $\xinner=0.5$, $\eta=0.2$, $L_{36} = 3$ (as in \POmain), $P = 1/550$~s and $R_6=1.12$. 
This value of the magnetic field is somewhat larger than it was estimated in other works. 
For example, \citet{zhang+1998b} estimated the magnetic field of the neutron star as $\sim 1\times 10^{8}$~G from the spectral analysis of 1997 outburst. 
The difference is caused by the  distance  they assumed ($2.5$~kpc; this explains factor of 2 since $B_x \propto L_x^{1/2}$) and their coefficient parametrizing inner radius corresponds to $\xinner\sim 1$:
 
 Similarly, \cite{asai+2013} and \cite{Campana+2014} suggested that the X-ray evolution during Aql\,X-1 outbursts of 1997 and 2010 demonstrated shut-off of the accretion process on to the neutron star surface due to the propeller effect. 
\cite{asai+2013} estimated the magnetic field as $(0.6-1.9) \times 10^8$~G analysing three outbursts from Aql\,X-1 from August 2009 to September 2012. 
The knee bolometric luminosity in our model \POmain is $\sim 3$ times more then they have estimated (partly because of the distance factor 2) and this explains our higher magnetic field estimate.
For the outbursts of 1997 and 2010, \cite{Campana+2014} estimated  the critical luminosity as $(5-6)\times 10^{36}$~erg/s and the magnetic field as $(1-4)\times10^8$~G for the distance 4.5~kpc.   
Evidently, the  crucial limitations to such estimates of the magnetic field are 
a rather uncertain factor $\xinner$ and details of the spectral modeling, which is rather simple in our study. 
Moreover, if the accretion inhibition works gradually,  such estimates are biased even more.

The concept of the disc evolution utilized in the present study has much similarity to a model suggested by~\citet{Hartman+2011}. 
They considered the double-flare of accretion-powered millisecond pulsar IGR J00291+5934 in 2008 and came to  a conclusion that (i) the disc had a shrinking zone of high viscosity; (ii) a knee on the light curves was likely to be due to the accretion inhibition by a propeller effect.  
We find that the X-ray evolution of Aql\,X-1 in 2013 can be explained satisfactorily  by the change of the hot zone size, but the additional increase in the  accretion efficiency when the disc edge is around $\Rcor$ makes the knee more pronounced.



\subsection{Persistent emission in the low state}\label{ss.plato}

The X-ray plato flux $2\times10^{-12}$ erg/cm$^2$/s obtained from the spectral modeling  (Fig.~\ref{fig.plato}) translates into $6\times 10^{33}$\, erg/s  in $0.5-10$~keV for isotropic radiation and 5~kpc. {\vsix Assuming a non-isotropic disc-like angular distribution of emission,  one obtains} $4 \times 10^{33}$\,erg/s in the same spectral  band (models \SARcor and \SARdead). Corresponding bolometric luminosities {\vsix of the disc} are $(1-2)\times 10^{34}$ erg/s (the interval appears due to different maximum temperature of the dead disc in \SARcor and \SARdead). Important to notice that the blackbody temperature of the neutron star surface would be about 0.2~keV,  consistent with observed estimates (see Fig.~\ref{fig.plato}).

If one explained the bolometric luminosity $1\times 10^{34}$ erg/s by the accretion onto the neutron star surface, that would demand the accretion rate $5.4 \times 10^{13}$~g/s or $8.5\times 10^{-13}\,\Msun$/yr. 
{\vsix This is much less than the transfer rate $\sim2\times10^{-10}~M_\odot$/year  suggested by \citet{Shahbaz+1998} in the framework of the DIM (see also refs.\ therein). 
There are two possibilities: the matter falls on the neutron star, but the rate is somehow restricted (while most matter is accumulated in the outer disc); or it is not the accretion onto the neutron star surface that produces the radiation.  }

 The model related to the first option is proposed by 
\citet{zhang+1998b, Shahbaz+1998, Menou+1999},  according to which the thin disc is substituted by an ADAF when the accretion rate is low, and the quasi-spherical flow overcomes the centrifugal barrier (if there is one, depending on the neutron star magnetic field strength), falls onto the neutron star, and gives  rise to the soft X-ray emission.

The X-ray radiation, {\vsix seen at the plato,} can be generated by the neutron star itself. 
A popular concept is that a cooling of an accretion-heated neutron star crust provides quiescent emission~\citep{Brown+1998,Wijnands+2017}.
X-ray radiation can also be produced far from the neutron star.
\citet{Cui1997} suggested for normal pulsars that X-ray emission in the low state is generated by relativistic particles accelerated in  a shock produced by the propeller wind colliding with the companion star wind.
When the stop radius for incoming matter  is larger than the light cylinder, the pulsar relativistic wind interacts with the matter inflowing through the Roche lobe, possibly giving rise to an X-ray emitting intrabinary shock ~\citep[see, e.g.,][]{Campana+1998,Bogdanov+2011,Wadiasingh+2018}.


{\vsix We  have considered an option that the radiation is produced mostly by the inner  remnant disc in scenarios \SARcor and \SARdead. }
In such a case, the X-ray plato slow evolution is determined by the  hot disc size evolution, which, in turn, is governed by the matter flow across the boundary between the hot and cold part of the disc.

\subsection{What makes different bursts: irradiation and peak accretion rate}
\begin{figure}
    \centering
    \includegraphics[width=0.47\textwidth]{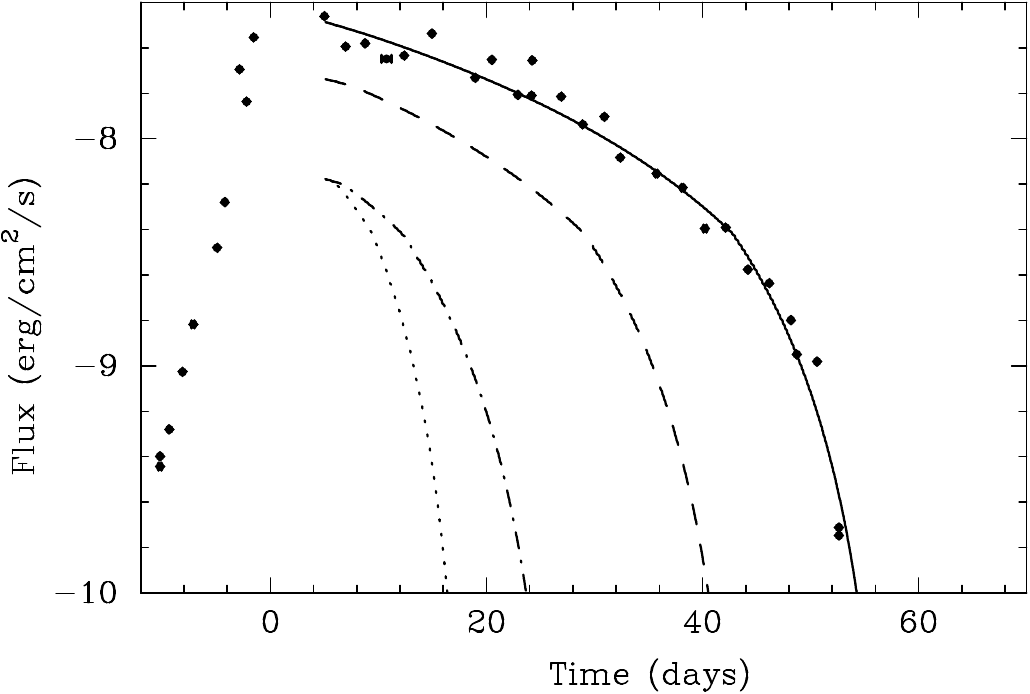}
     \includegraphics[width=0.47\textwidth]{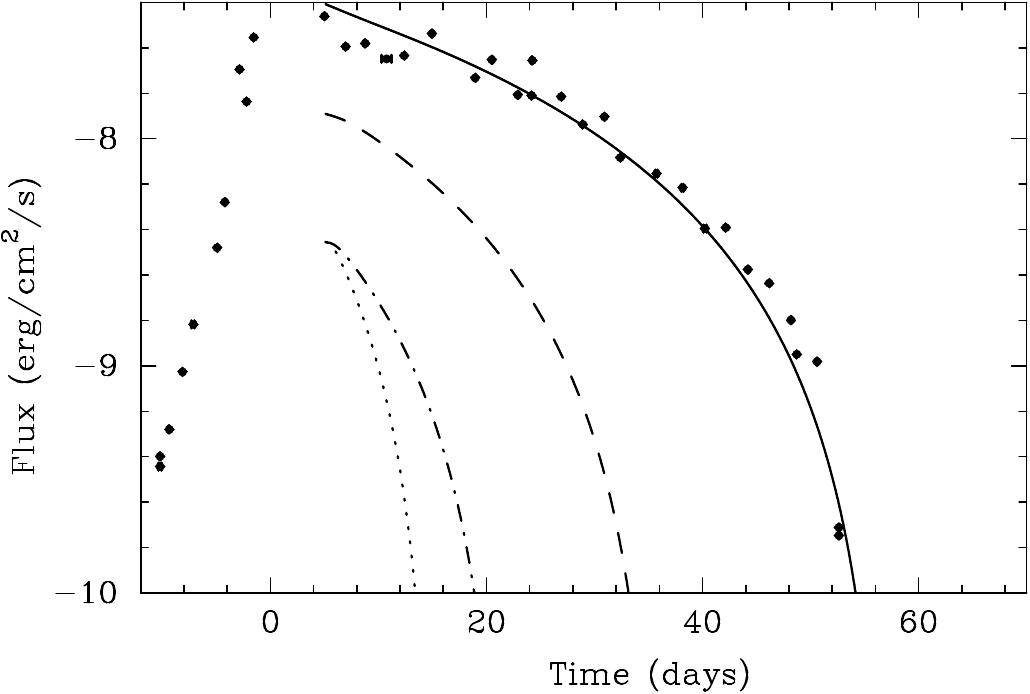}
      \caption{Upper panel: Modeled light curves with different peak accretion rates $\dot{M_0}$ ($7.3\times 10^{17}$, $3.5\times 10^{17}$, and $1\times 10^{17}$~g/s), $B_x\approx 6.3\times 10^{8}$~G, $\alpha = 0.74$, and $\Chot=0.0014$.  The lowest   dotted model curve with  faster decay has $\Chot=0.0007$. The  2013 outburst is shown with dots.In the lower panel the light curves are calculated for the same disc parameters, except that $\dot{M_0}=1.1\times 10^{18}$, $3.5\times 10^{17}$, and $1\times 10^{17}$~g/s and the magnetic field $B_x=100$~G.}
    \label{fig.Rhot_Mdot}
\end{figure}

Using our code, we can generate a family of  light curves of a source, varying only the peak accretion rate. Figure~\ref{fig.Rhot_Mdot} shows such X-ray light curves for the same parameters as we use for Aql\,X-1. All curves, except one, have the same $\alpha$ and $\mathcal{C}_\irr$ (see the caption). The lowest curve is calculated for a smaller value of $\mathcal{C}_{\rm irr}$; subsequently, the radius of the hot disc is smaller, and the evolution is faster. 

The lower panel of Fig.~\ref{fig.Rhot_Mdot} presents a family of light curves of  an X-ray transient with a negligible magnetic field of neutron star and a shrinking hot zone. 
The fast evolution can be seen, although a characteristic knee is smoothed out. 
The knee is more pronounced in the models with high magnetic field  because the accretion efficiency in the disc {\vsix varies non-monotonically} (see Fig.~\ref{fig.aux_pars}, the blue dashed line). 
When the magnetic field is low, the accretion rate efficiency is constant~(see model \smallB in Supplement Material). 
Less width of outbursts  is explained by smaller hot zones due to  lower X-ray emission that lacks the energy gained from the rotation of a fast-rotating magnetized neutron star (cf. Eq.\ref{eq.L_bolometric}). Its contribution is most notable when the inner disc radius approaches the corotation radius.

\section{Summary}\label{s.summmary}
A model and an computer code \freddi are presented to calculate viscous disc evolution in an X-ray transient with an  accreting magnetized neutron star. 
The accretion rate  evolution is obtained by solving the viscous evolution equation for an $\alpha$--disc, with no {\em ad hoc} time dependences. 
The key parameters of the model are  the mass, spin frequency, and magnetic field of the accretor, the $\alpha$-parameter and the degree of self-irradiation of the disc.
Light curves produced by  thermal radiation  can be calculated with \freddi.

Specific boundary conditions are set in order to input essential physics of the disc-magnetosphere interactions. 
Our method uses an inference that viscous and magnetohydrodinamic  processes at the disc inner boundary, being relatively fast on a whole-disc timescale,  can be time-averaged and approximated by smooth conditions when solving the equation of the  disc evolution.

Three  scenarios of centrifugal inhibition of aaccretion on a neutron star are included so far: `instant' block of accretion and effective propeller outflow  from the corotation radius (\PO), {\vsix MHD-calculations-based} gradual blocking and propelling with continued star accretion (\PONS), and no mass propelling (\NS). 

We apply our  model to an outburst of Aql\,X-1. 
This source and this particular outburst of 2013 have got a wealth of observational data in different spectral bands. 
The model is successful at explaining the X-ray and optical evolution, including a characteristic fast drop of the X-ray light curve.
                
The irradiation of the disc plays a major role 
in the evolution of the  outburst of  Aql\,X-1. 
In particular, the characteristic bend (a `knee') of the X-ray light curve is explained  by the viscous evolution of the hot zone with time-dependent outer radius determined by the condition of the irradiation temperature, $T_\irr=10^4$~K. 
Resulting parameter $\alpha$ is positively correlated  with the irradiation parameter $\Chot$. 
This is due to the fact that increasing turbulent parameter $\alpha$ (or kinematic coefficient of viscosity $\nu_t$) accelerates the mass-accretion evolution, and increasing  the irradiation parameter $\Chot$ (or the size of the ionized disc) decelerates the evolution since characteristic time $\propto R_{\rm hot}^2/\nu_t$. 
In the best models  the hot part of the disc (where $T_\irr>10^4$~K) can  extend to the radius $0.85-0.95$ of $\sim R_{\rm tid}$ at the peak of the outburst. 
This radius correlates with the irradiation parameter, which is estimated as  $\Cirr^{\rm max} = (1.4-2.8)\cdot 10^{-3}$  for $\alpha=0.5-0.8$.

A variation of $\Cirr$ with time is strongly favoured in the model to satisfactory explain the form of the X-ray light curve. 
This is consistent with being the geometry effect.

Assessment of $\alpha$ requires a satisfactory fit to optical, which is hampered by uncertainties of the distance to the source, optical star temperature, etc.
For the parameters assumed, the optical flux  is  explained by the irradiation of the disc with the geometrical size close to the tidal radius  $\sim R_{\rm tid}$ with a contribution from the optical companion.
A scatter of the optical data indicate that  reprocessing of X-rays occurs with an orbit- or near-orbital modulation, superimposed on the X-ray variations, seen  on a scale of days.   
We find that the optical data scatter is comparable with the oscillations produced by the optical companion with albedo $\gtrsim0.5$. 
Ignoring optical contribution the cold disc and  optical companion results in a larger hot disc, which naturally affects  the rate of the viscous evolution, requiring larger $\alpha$.


We find that the  light curve of Aql\, X-1\,(2013) can be tolerably approximated without invoking the magnetosphere action  (with coefficient of determination $\mathcal{R}^2\approx 0.98$). 
At the same time, models with magnetic field $>10^8$~G are statistically preferred ($\chi^2$ is 1.5 times less) because they produce a more pronounced knee on a light curve due to the accretion-efficiency dependence on the inner disc radius.
For high magnetic fields, $>10^8$~G, we find that the X-ray data favors the scenarios with gradual extinction of matter flow to the neutron star surface. Scenario \PO with an abrupt block of accretion flow at the centrifugal barrier does twice as worse comparing to scenarios with a continued accretion through the magnetospheric boundary. 
Resulting values of the magnetic field are $(5-9) \cdot 10^8$~G  for \NS and \PONS scenarios and $(3-4) \cdot 10^8$~G, for the  \PO scenario.  
These values depend strongly on the definition of the magnetospheric radius \eqref{eq.rmag} and distance to the source. 

We also find an interesting possibility that the X-ray plato is produced by a remnant disc-reservoir with stalled accretion. This is possible if the magnetic field is in the range $3\times 10^7 - 6\times 10^8$~G (for distance 5 kpc). 
The temperature $0.2-0.3$~keV of the blackbody component in the X-ray `blackbody plus power-law tail' spectrum of Aql\,X-1 in 2013 is about the maximum temperature of that disc-reservoir.

\section*{Acknowledgments}

The authors are grateful to the anonymous referee for extensive helpful
suggestions and to Alexander Meshcheryakov for the fruitful discussions. 
This work is supported by the Russian Science Foundation grant 21-12-00141
and performed in Moscow Lomonosov State University (theoretical model
development and model analysis of the Aql X-1 2013 outburst).  
Optical data processing was supported by the grant 14.W03.31.0021 of the Ministry of Science and Higher Education of the Russian Federation and the grant from Academy of Finland 332666.
%
X-ray data spectral
modeling was supported by the Interdisciplinary Scientific and Educational
School of Moscow University "Fundamental and Applied Space Research".  This
research has made use of data obtained through the High Energy Astrophysics
Science Archive Research Center Online Service, provided by the NASA/Goddard
Space Flight Center.  This research has made use of the NASA/IPAC Infrared
Science Archive, which is funded by the National Aeronautics and Space
Administration and operated by the California Institute of Technology. This
work made use of data supplied by the UK Swift Science Data Centre at the
University of Leicester.  This paper has made use of the SMARTS
optical/near-infrared light curves.  The Yale SMARTS XRB team is supported
by NSF grants 0407063 and 070707.  This research has made use of NASA’s
Astrophysics Data System Bibliographic Services.

 \section*{Data  availability}
Authors intend to make available unabsorbed X-ray flux in 0.5-10~keV and optical flux density data of Aql\,X-1 used for fitting at VizieR database of astronomical catalogues. Code \freddi can be found at \url{https://ascl.net/1610.014}.
 
\bibliographystyle{mnras}
\bibliography{burst_evol_Aql}

\appendix

\section{Vertical structure}
\label{Appendix.OPAL_vertical_structure}

\begin{figure}
    \centering
    \includegraphics[width=0.46\textwidth]{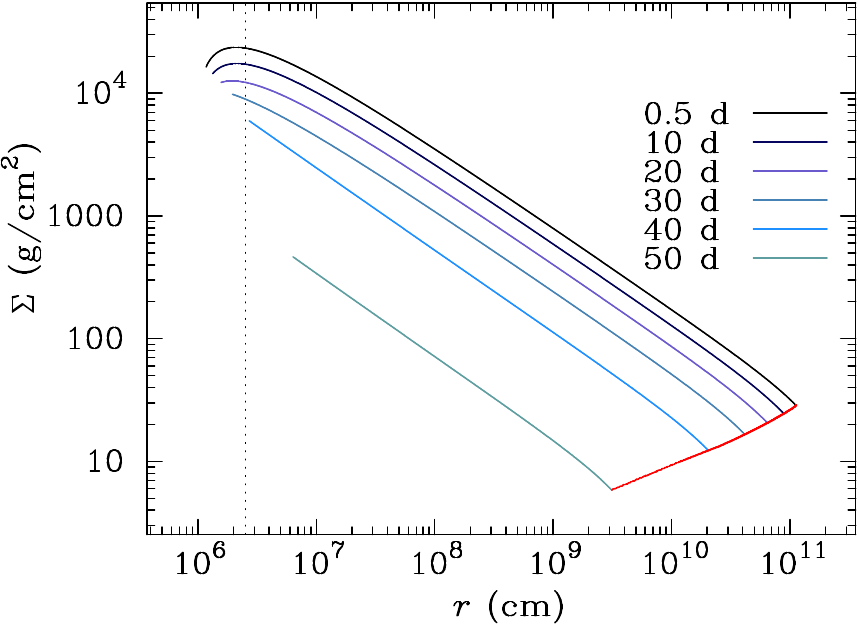}
    \caption{Surface density distributions in the ionized part of the disc in \SAmain model  at different times labeled from top to bottom. The vertical line marks the corotation radius. 
    }
    \label{fig.disk_Sigma}
\end{figure}
To obtain a relation between the surface density $\Sigma$ and the viscous stress tensor $W_{r\varphi}$, one needs to solve the vertical structure equations. 
In the {\freddi} code, physical parameters (column density $\Sigma$, semithickness $z_0$, etc.) at any radius are found using analytic approximations to numerical solutions of the disc vertical structure~\citep{ketsaris_shakura1998,malanchev_et2017,Lipunova+2018chapter}. 
The surface density in the disc is as follows:
$$
\Sigma \approx 34\, \mbox{g/cm}^2 \,\alpha^{-7/9}\, r_{10}^{-2/3}\, \dot M_{17}^{2/3} \, m_x^{2/9}  \,
$$
for the opacity  approximation of \citet{bell-lin1994},  appropriate for the solar chemical abundances:
$$
\varkappa =\varkappa_0\, \frac{\rho^\varsigma}{T^\curlyvee}\, ,\quad
\varsigma\approx 1, \curlyvee\approx 5/2, \quad
\varkappa_0\approx 
1.5\cdot 10^{20}~\mbox{cm}^5\mbox{K}^{5/2}\mbox{g}^{-2}\, .
$$
In Fig.~\ref{fig.disk_Sigma}, the surface density distributions are presented, corresponding to torque distributions shown in Fig.~\ref{fig:freddi_F_h} in model \SAmain.
The red line is the surface density left after the burst passed by, $\Sigma(\Rhot(t))$. 
Qualitatively, this picture resembles the burst evolution obtained by \citet{dubus_et2001} (see their figure 12).
Corresponding disc semithickness is
$$
z_0/r \approx 0.05\,  \alpha^{-1/9}\, r_{10}^{1/12}\, \dot M_{17}^{1/6} \, m_x^{-13/36}  \, .
$$

We note that a structure of a non-irradiated disc is used in the model with irradiation contributing significantly to the optical flux, which is justified by the previously obtained results. 
As suggested by \citet{Lyutyi+Sunyaev1976} and confirmed by  \citet{mescheryakov_etal2011}, the irradiated disc in the hot zone does not change significantly its vertical structure comparing to the case without irradiation, due to its large optical thickness (see also \citealt{dubus_et1999}).

The region, evacuated by the hot disc when it shrinks, is likely optically-thin. 
In such a case, to explain the optical `black-body' emission from a cold zone of size of the tidal radius $\sim R_{\rm tid}$ we suggest that there is a `cold' dense ring of non-ionised matter near  $R_{\rm tid}$ that survives an outburst. 
Its temperature is very low, and we assume that it shines only by reprocessing X-rays, possibly scattered by the plasma above the disc.
This ring casts a shadow on the optical companion {\vsix when it is thicker than the hot disc}.     
The optically thin matter may form a corona, similar to the Solar one, with characteristic temperatures around $10^6$~K at distances $\sim 10^{11}$~cm. {\vfour The characteristic time of such corona formation is $1/(\alpha\, \omega_{\rm K})$.}

\begin{figure*}
    \centering
    \includegraphics[width=0.95\textwidth]{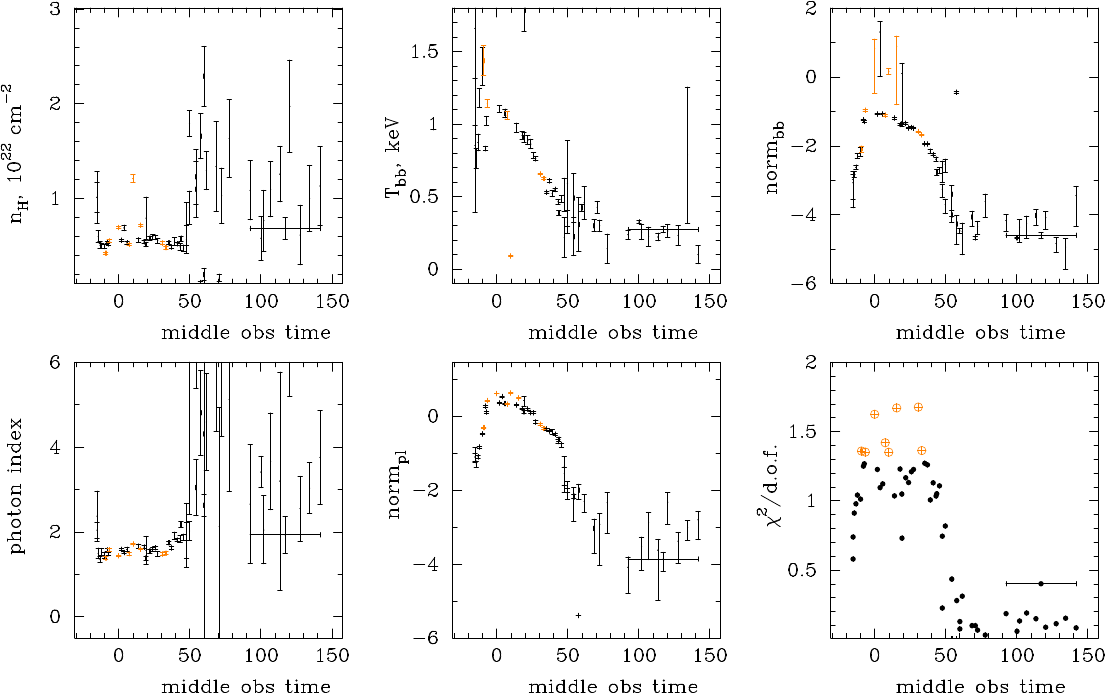}
    \caption{Evolution of the spectral parameters for model {\tt tbabs*(bbody+powerlaw)}. Orange color indicate fits with reduced $\chi^2>1.3$. Zero time is  MJD~56466.763. {\vfour The point with the long horizontal bar is derived from the spectrum integrated over MJD\,56550-56650 and used for fits \SARcor and \SARdead.}
    }
     \label{fig:sppar_evolution}
\end{figure*}

\section{Bolometric luminosity of the neutron star}\label{s.bolLns}
The momentum equation for the matter moving in the magnetosphere yields the following energy integral~\mbox{\citep{ghosh_etal1977i,Ustyugova+1999}}:
$$
\frac 12\, (v_p^2 + \omega^2\, r^2) - \omega_\star\, \omega\, r^2 - \frac{G\, M_\star}{r} = const\, ,
$$
where $v_p$ is the poloidal speed and $\omega$ is the angular velocity of the matter. We assume that the poloidal speed at inner disc radius $\Rin$ is zero and the radiated energy results from the kinetic energy of the poloidal speed at the neutron star surface.  Then, luminosity due to accretion onto a neutron star can be written as follows
\begin{equation}
\begin{split}
L_\star = \frac{\dot M_\star \, G\, \mns}{R_\star} \, \left(1 - \frac{R_\star}{\Rin}\right) + \frac{\dot M_\star\,\omega_\star^2}{2}\,(R_\star^2-\Rin^2)
 \\
 + \frac{\dot M_{\star} \, G\, M_{\star}}{2\, \Rin} \left(1-\frac{\omega_{\star}}{\omega_{K(\Rin)}}\right)^2\, . 
 \end{split}
\label{eq.Lstar}
\end{equation}
and  \eqref{eq.general_eta_ns} is easily obtained.
The terms in the last expression were rearranged in an attempt to associate them with specific processes during the matter motion.
The first term is the work by the gravitational forces. 
The second term is the change of the `centrifugal energy' in the co-rotating coordinate system (c.f. \citealt[\S39,][]{landau1976mechanics}), or the work by electromagnetic forces in the inertial coordinate system.  
The third term is due to the braking of the matter angular velocity from the Keplerian  ${\omega_K}(\Rin)$ to the magnetospheric ${\omega}_{\star}$. 
This term can be compared to the heat released at the surface of a star rotating slower than an adjacent disc, which is obtained from the conservation of momentum and energy~\citep[e.g.,][]{Shakura-Sunyaev1988,kley1991}: 
\begin{equation}
L_\star   = \frac{\dot M \, G\, \mns} {2\, R_\star} \left(1- \frac{\omega_\star}{\omega_K(R_\star)}\right)^2\, .
\label{eq.eff_kley}
\end{equation}
 When the magnetic field is not important, the orbiting matter decelerates from a Keplerian  velocity  to  the  rotational  velocity  on  the  neutron star, and the extra energy is turned into heat. 

Expression \eqref{eq.Lstar} applies for both the case of small and large magnetic field. 
It depends on the accretion rate, if there is a magnetosphere ($\Rm>R_\star$).  
Some of the kinetic energy, related to the orbital motion of the matter in the disc is spent on the neutron-star's spin-up, and the rest energy is released as the heat on the neutron star's surface.
For high accretion rates, the disc overwhelms the magnetic field, which can be thus ignored. 
Then $\Rm = R_\star$, and only the last term survives in \eqref{eq.Lstar}, becoming \eqref{eq.eff_kley}, independent from $\dot M$. 
If $\Rm = R_\star$ and the star does not rotate, it radiates the same energy as the disc.

\section{Optical extinction towards Aql\,X-1}\label{sss.redening}
IRSA Dust service\footnote{\url{https://irsa.ipac.caltech.edu/applications/DUST/}} gives colour excess in the direction the source is $E_{B-V} = 0.61\pm0.02$ for COBE/DIRBE and IRAS/ISSA maps \citep{Schlegel_etal1998} and $E_{B-V} = 0.71\pm0.03$ for \citet{Schlafly-Finkbeiner2011} map based on SDSS spectral data.
3D Dust Mapping service\footnote{\url{http://argonaut.skymaps.info/query}} gives colour excess $E_{B-V} = 0.64^{+0.02}_{-0.03}$ for distance of 5\,kpc~\citep{Green2019}.
According to these data we use 
$E_{B-V} = 0.64 \pm 0.04$, taking into account  the uncertainty in  the photometric data.
We use standard relation between the colour excess and visual extinction: $A_V = 3.1\, E_{B_V}$ \citep{Schultz_Wiemer1975}.

\citet{Sakurai+2012}, 
performing for the  Aql\,X-1 outburst of 2007  spectral fits in 0.5$-$30~keV, obtains $N_H = 3.6\cdot10^{21}\,\mathrm{atoms}/\mathrm{cm}^2$.
Taking into account $N_H / E_{B-V} = 5.8\cdot10^{21}\,\mathrm{atoms}/\mathrm{cm}^2/\mathrm{mag}$ \citep{Bohlin_etal1978} it would give $E_{B-V} = 0.62$.
Our spectral modeling of the Swift/XRT data gives $N_H= (5.5\pm0.2)\cdot10^{21}\,\mathrm{atoms}/\mathrm{cm}^2$ (see \S\ref{ss.XRT}).
Thus, the  values obtained from X-ray observations,  agree with the above $E(B-V)$.

We transform the visual extinction to {\vsix that} in particular optical passbands using SVO Filter Profile Service\footnote{\url{http://svo2.cab.inta-csic.es/theory/fps/}} \citep{Rodrigo_etal2012}.
It gives the following values for the $A_f / A_V$ ratio: $UVW2=3.09$, $UVM2=3.02$, $UVW1=2.06$, $U=1.62$, $B=1.32$, $V=1.02$, $R=0.82$, and $J=0.3$.
\label{lastpage}

\end{document}